\documentclass{article}

\usepackage[preprint]{neurips_2026}

\usepackage[utf8]{inputenc} 
\usepackage[T1]{fontenc}    
\usepackage{hyperref}       
\usepackage{url}            
\usepackage{booktabs}       
\usepackage{amsfonts}       
\usepackage{nicefrac}       
\usepackage{microtype}      
\usepackage{xcolor}         
\usepackage{amsmath}
\usepackage{graphicx}
\usepackage{sidecap}
\sidecaptionvpos{figure}{c}

\title{Calibrating subgrid parametrizations of single-column
ocean models via simulation-based inference}
\workshoptitle{Sim2Science}

\author{%
  Luben M. C. Cabezas \\
  DEs, ICMC\\
  Federal Univ. of São Carlos, Univ. of São Paulo, Brazil \\
  Univ. Grenoble Alpes, Inria, \\
  CNRS, Grenoble INP, LJK, France \\
  \texttt{lucruz45.cab@gmail.com} \\
  \And
  Sacha Wendling \\
Univ. Grenoble Alpes, Inria, \\
CNRS, Grenoble INP, LIG, France \\
  \texttt{sacha.wendling@inria.fr} \\
  \And
  Aurèle Gallard \\
Univ. Grenoble Alpes, Inria, \\
CNRS, Grenoble INP, LJK, France \\
  \texttt{aurele.gallard@inria.fr} \\    
  \And
  Gabriel Mouttapa \\
Univ. Grenoble Alpes, Inria, \\
CNRS, IRD, Grenoble INP, INRAE, IGE, France \\
  \texttt{gabriel.mouttapa@univ-grenoble-alpes.fr} \\   
  \And
  Julien Le Sommer \\
Univ. Grenoble Alpes, Inria, \\
CNRS, IRD, Grenoble INP, INRAE, IGE, France \\
  \texttt{julien.le-sommer@univ-grenoble-alpes.fr} \\   
  \And  
  Pedro L. C. Rodrigues \\
Univ. Grenoble Alpes, Inria, \\
CNRS, Grenoble INP, LJK, France \\
  \texttt{pedro.rodrigues@inria.fr} \\  
}

\begin{document}

\maketitle

\begin{abstract}
Subgrid parametrizations of vertical mixing in ocean models depend on free
coefficients that cannot be measured directly and must be calibrated against
high-fidelity references such as large-eddy simulations (LES). Existing
approaches return point estimates and leave the associated uncertainty unquantified, a
limitation when the inverse problem is ill-posed or when distinct parameter
configurations fit the data comparably well. Simulation-based inference (SBI)
addresses exactly this: given a prior and access to the simulator, it
approximates the full posterior over parameters without requiring a tractable
likelihood, at a cost set by the number of simulator evaluations. We apply it to
\texttt{tunax}, a JAX-based single-column ocean model, to calibrate the
coefficients of its $k$--$\varepsilon$ closure. A blockwise PCA summary statistic
compresses the simulator output along the depth axis while preserving its
forcing--horizon--variable structure, making inference tractable at modest
budgets. We compare neural posterior estimation and its sequential variants
against a recent training-free approach built on a tabular foundation model. The
latter recovers informative posteriors from a few hundred simulator calls,
outperforming the trained estimators at every budget considered.
\end{abstract}

\section{Introduction}

Ocean and climate models run on grids far too coarse to resolve the turbulent eddies that actually mix heat, momentum and tracers through the water column. Since simulating that turbulence explicitly is computationally out of reach, its net effect is instead \emph{parametrized}: replaced by closed-form equations predicting the mixing from the large-scale state at negligible cost, at the price of free coefficients $\boldsymbol{\theta}$ that are not directly measurable and must be \emph{calibrated} \citep{schneider2017earth}.

Single-column ocean models (SCMs) are the natural testbed for such schemes: they retain the vertical structure of a water column and the parametrized mixing acting on it, while ignoring horizontal dynamics~\citep{hartung2018ec}. Despite this simplicity, they are a cornerstone of climate modelling — the vertical mixing schemes they embody are directly inherited by general circulation models (GCMs), where the same closure operates column by column~\citep{reichl2024improving} — so errors in tuning propagate to large-scale predictions of heat transport, stratification, and sea surface temperature. A common way to calibrate SCMs is to target Large-Eddy Simulation (LES) data. LES resolves explicitly, at fine spatial scales, the very eddies the parametrization stands in for, making it a high-fidelity but computationally expensive reference.
Given a set of LES
outputs $\mathbf{x}_\mathrm{LES}$ obtained under specific atmospheric forcings,
the calibration problem is to find parameters $\boldsymbol{\theta}$ such that
the corresponding output of the SCM closely matches
$\mathbf{x}_\mathrm{LES}$. 

\texttt{tunax}~\citep{tunaxurl} is a modern JAX-based implementation of a
single-column ocean model. It simulates the evolution of temperature and
velocity profiles under prescribed atmospheric forcings, capturing boundary-layer
turbulence, convective mixing, and shear-driven transport. Crucially, \texttt{tunax} simulations are computationally inexpensive — a single
forward run takes only a few seconds on a standard CPU — making it an ideal
testbed for benchmarking and comparing different calibration strategies.

Existing approaches to SCM calibration typically yield point estimates of
$\boldsymbol{\theta}$: parameters are either tuned by hand~\citep{reffray2015modelling}, optimized via gradient-based algorithms when the simulator is
differentiable \cite{aldebert2021fast}, or inferred through ensemble methods such as Ensemble
Kalman Inversion (EKI) and its variants~\citep{Iglesias_2013, Calvello_Reich_Stuart_2025, wagner2025formulation, gjini2025ensemble}. While effective, these approaches do not quantify uncertainty over
parameters — a significant limitation when the calibration problem is
ill-posed or when multiple parameter configurations produce similarly good fits
to the LES data.


Simulation-based inference (SBI)~\citep{deistler2025simulation} offers a principled alternative: given a
prior $p(\boldsymbol{\theta})$ and access to the simulator, it approximates
the full posterior $p(\boldsymbol{\theta} \mid \mathbf{x}_\mathrm{LES})$
without requiring an explicit likelihood. The low cost of \texttt{tunax}
simulations makes this setting particularly amenable to SBI, where the
inference quality scales with the number of simulator evaluations. In this
work, we compare two SBI approaches: neural posterior estimation (NPE)
\citep{cranmer2020frontier}, which trains a normalizing flow on simulated
$(\boldsymbol{\theta}, \mathbf{x})$ pairs, and the recent TabPFN-based method
of~\citep{vetter2025effortless}, which leverages a pre-trained in-context model to
approximate the posterior from very few simulations without any additional
training.

We demonstrate, to our knowledge, the first application of SBI to the
calibration of ocean climate models, using \texttt{tunax} as a
testbed. Our contributions are: (i)~a PCA-based summary statistic
construction tailored to the spatiotemporal structure of SCM outputs;
(ii)~a comparison of NPE and TabPFN-SBI across varying simulation budgets,
evaluated on real LES outputs; and (iii)~evidence that SBI is a viable strategy for calibrating \texttt{tunax} parameters, opening a path toward
uncertainty-aware calibration of ocean subgrid parametrizations.

\section{Background}
\label{sec:background}

Let $p(\boldsymbol{\theta})$ be a prior over parameters and
$p(\mathbf{x} \mid \boldsymbol{\theta})$ a simulator-induced likelihood that
is intractable to evaluate but possible to sample from. SBI methods approximate the posterior
$p(\boldsymbol{\theta} \mid \mathbf{x})$ using a dataset of simulated pairs
$\{(\boldsymbol{\theta}^{(i)}, \mathbf{x}^{(i)})\}_{i=1}^N$.
Neural Posterior Estimation (NPE) instantiates this by training a
conditional density estimator $q_\phi(\boldsymbol{\theta} \mid \mathbf{x})$
— typically a normalizing flow \citep{papamakarios2021nf} — to approximate the posterior directly.
A key design choice is the proposal distribution from which simulations are
drawn: \emph{amortized} NPE samples from the prior $p(\boldsymbol{\theta})$,
producing a network valid for any $\mathbf{x}$, while \emph{sequential} NPE~\citep{greenberg2019automatic}
refines the proposal across multiple rounds to focus simulations on the
posterior for one specific observation $\mathbf{x}_o$.

In the applied setting considered in this work, amortization offers little benefit: we target a single fixed
LES run, and there is no need for a general-purpose amortized posterior network. The
sequential approach is therefore more natural, as it concentrates the
simulation budget in regions of parameter space consistent with
$\mathbf{x}_\mathrm{LES}$, yielding a better-resolved posterior for the same
number of simulator calls. A recent alternative is the TabPFN-based approach of~\citet{vetter2025effortless}, which
leverages the in-context learning capabilities of a pre-trained tabular foundational model to approximate the posterior from very few
simulations without fitting any additional parameters. Despite being amortized
in its training, it achieves competitive accuracy at low simulation budgets,
making it an attractive complement to SNPE-C in budget-constrained settings
such as ours.

\texttt{Tunax} instantiates the simulator
$\mathbf{x} \sim p(\mathbf{x} \mid \boldsymbol\theta)$ above; we
calibrate its 17 coefficients, governing the production and dissipation
of turbulent kinetic energy and its coupling to buoyancy-driven mixing. More details on the coefficients can be found on Appendix \ref{app:prior_setup}. Each forward run outputs profiles of buoyancy, the two horizontal
velocity components, and a passive tracer — four variables in total —
under $7$ forcing regimes and $3$ forecast horizons drawn from the LES
reference dataset \citep{wagner2025formulation, wagner2026reference}.

\section{Experimental setup}
\label{sec:setup}

\textbf{Simulations setup.}
We use a log-normal prior over parameters $\boldsymbol{\theta}$ — equivalently, a
normal prior with standard deviation $\log(1.5)$ on the log-transformed
parameters, with mean equal to the log-transformed literature default
values \citep{Umlauf2005} — giving roughly a $1.5\times$ multiplicative range
around each default. Two coefficients whose physical role requires a
fixed sign are log-transformed with that sign fixed accordingly; details
on the prior, the default values and these sign-constrained parameters are given in
Appendix~\ref{app:prior_setup}. The raw simulator output used here is a set of depth profiles at
$4\,\mathrm{m}$ vertical resolution, obtained by retaining only the last state of each simulated time series, one per (forcing, time horizon, variable) combination---$7 \times 3 \times 4 = 84$ profiles, each over a fixed 54-level water-column window. Flattening the full output ($D_{\mathrm{raw}}=4536$) and masking the
uninformative forcing--variable combinations described in
Appendix~\ref{app:prior_setup} yields the retained observation vector
$\mathbf{x}\in\mathbb{R}^{D}$ with $D=3888$, later reduced via the
summary statistic in Section~\ref{sec:setup}. While the underlying turbulence is stochastic, \texttt{tunax} implements
it as a deterministic mapping $\boldsymbol\theta \mapsto \mathbf{x}$,
which alone induces a degenerate likelihood unsuitable for SBI. We
therefore add Gaussian observation noise, a standard device for
approximating simulator discrepancies as Gaussian~\citep{wood2010statistical,Iglesias_2013}; concretely, even at
the parameters that best explain $\mathbf{x}_\mathrm{LES}$, this
reduced-physics single-column model cannot be expected to match the
higher-fidelity LES reference exactly, so the noise also captures this
model discrepancy~\citep{kennedy2001bayesian}. Its covariance is
estimated empirically from the same prior predictive pilot sample used
to fit the PCA summary statistics (Appendix~\ref{app:noise}).

\textbf{Dimensionality reduction via PCA.}
Flattening $\mathbf{x} \in \mathbb{R}^D$ into a single vector does not,
on its own, resolve its dimensionality for simulation-based inference —
it also discards the block structure above, treating physically distinct
regimes (e.g.\ the depth profile under free convection versus under
strong wind forcing) as homogeneous coordinates of one measurement. We instead reduce dimensionality along the depth axis independently within each
(forcing, horizon, variable) block. We use the outputs of $N=1000$ prior
predictive simulations to fit a separate PCA basis for each block
\citep{blum2013comparative,jolliffe2016principal}, retaining the minimum number of components
required to explain $90 \%$ of the block's variance, up to
$k_{\max}=5$. The selected dimensions and explained variances are reported in
Appendix~\ref{app:pca}. The retained coefficients are concatenated into the
summary vector $\mathbf{s}(\mathbf{x})\in\mathbb{R}^{d_s}$, with
$d_s=159\ll D$, which serves as the observation operator throughout the
calibration. Gaussian noise is subsequently added to these coefficients to
obtain the stochastic simulator output used for inference (Appendix \ref{app:noise}).

\textbf{Inference methods and evaluation protocol.}
We compare two families of methods: neural density estimators trained
from scratch — amortized NPE and its two sequential variants,
SNPE~\citep{greenberg2019automatic} and
TSNPE~\citep{deistler2022truncated}, which update the proposal
differently across rounds — and the training-free, TabPFN-based
NPE-PFN~\citep{vetter2025effortless}, a pre-trained foundation model
applied in-context to $(\boldsymbol{\theta}, \mathbf{s}(\mathbf{x}))$
examples. We are interested both in whether sequential refinement
improves over amortized training within the first family, and in how
NPE-PFN compares to it — whether its training-free posterior is
especially advantageous at low simulation budgets, and whether that
advantage persists as the budget grows. We sweep the simulation budget between $50$ and $3000$ (details in Appendix \ref{app:additional_results})
simulator calls to answer both questions, checking whether small budgets
already match larger ones or whether more simulations keep paying off.
For each method, let $\bar{\boldsymbol\theta} = \mathbb{E}[\boldsymbol\theta
\mid \mathbf{s}(\mathbf{x}_\mathrm{LES})]$ denote its posterior mean. We
compare the \texttt{tunax} trajectory at $\bar{\boldsymbol\theta}$
against the real LES trajectory $\mathbf{x}_\mathrm{LES}$, and
$\bar{\boldsymbol\theta}$ against a gradient-based calibration estimate
obtained via JAX automatic differentiation $\boldsymbol\theta_{\mathrm{GD}}$, with the full comparison
regarding this calibration reported in Appendix~\ref{app:gradient_based_results}. Although this procedure provides only a point estimate, we use it as a proxy for the unknown effective parameters that best reproduce the LES within \texttt{tunax}. The distance between a method's posterior mean and this proxy measures whether its central location approaches the gradient-calibrated solution, while the density comparison assesses whether the proxy lies within regions of high posterior density.

\section{Numerical illustrations}

\textbf{Accuracy versus simulation budget.}
We first compare four inference methods across simulation budgets in output
space. For each method and budget, we run \texttt{tunax} at the posterior mean
$\bar{\boldsymbol{\theta}}$ and report the profile-standardized RMSE between the
resulting profiles and the LES reference (Figure~1). All methods improve
substantially over the default \texttt{tunax} configuration, and all of them
plateau: because \texttt{tunax} is a reduced-physics model, no parameter value
reproduces the LES exactly, so the achievable RMSE is bounded away from zero and
the relevant question is how quickly, and how closely, each method approaches
that bound. NPE-PFN descends fastest --- it is already near its plateau at a few
hundred simulations --- and remains the best posterior method from moderate
budgets onward. The sequential variants SNPE and TSNPE improve over amortized
NPE, as expected when the simulation budget is concentrated on a single
observation, but neither matches NPE-PFN at any budget considered here.

As a reference point, we also report a gradient-based calibration
(\textsc{grad-full-profiles}): exploiting the differentiability of the JAX
implementation, we directly minimize the discrepancy to the LES profiles by
gradient descent, obtaining the point estimate $\boldsymbol{\theta}_{\mathrm{GD}}$.
This is the natural approach when it is available, and it attains a visibly lower
RMSE than any posterior mean. Two caveats temper the comparison. First, it
requires a differentiable simulator, which most ocean and climate codes are not, which is precisely the setting SBI is designed for. Second, its cost is not
comparable to a forward call: the marker is placed at $10^3$ simulator calls, but
each of those calls is a differentiated one and therefore substantially more
expensive than an ordinary forward simulation. Finally, it returns a single
point, with no account of which other parameter values would explain the data
comparably well.

\begin{SCfigure}[1.0][t]
    \centering
    \includegraphics[width=0.5\linewidth]{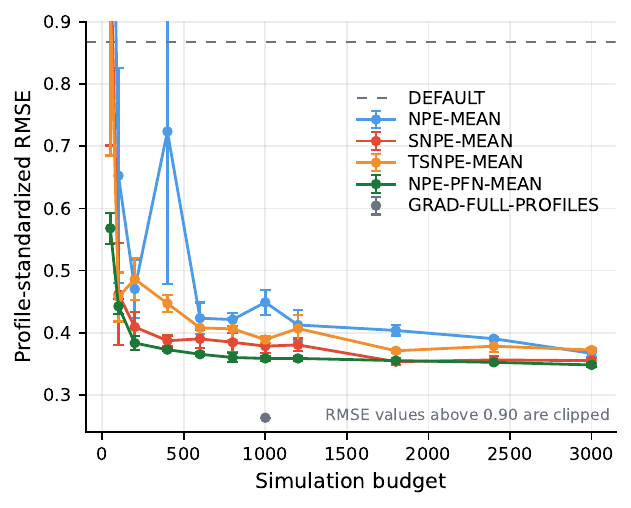}
    \caption{
RMSE between LES profiles and the \texttt{tunax}
output evaluated at each posterior mean. Markers and error bars show the mean and SEM across 5 realizations, and the dashed line denotes the default \texttt{tunax} configuration. The gray marker denotes full-profile gradient calibration; its position counts $1000$ simulator calls, although differentiation makes these calls more costly than ordinary forward simulations. NPE-PFN-MEAN reaches lower RMSE rapidly at small budgets and performs best among the posterior methods from moderate budgets onward. All methods plateau at budgets of around 1000 simulations, yet none approach the gradient descent optimum.
}
    \label{fig:profile_standardized_rmse_vs_budget}
\end{SCfigure}

\textbf{Prior and posterior predictive checks.} Figure 2 examines the performance of the methods in output space, contrasting the posterior predictive distributions against the prior predictive as a no-inference baseline. Under strong-wind forcing at 6 hours,
we compare prior predictive profiles against the posterior predictives of SNPE
and NPE-PFN for $b$, $u$, $v$ and $p_t$. $b$ is the buoyancy of the water, it's a proxy of the temperature, $u$ and $v$ are the horizontal velocities of the water and $p_t$ is a passive tracer (a dimensionless variable that is only diffused by the turbulence). SNPE shifts the predictive medians
toward the LES curves but leaves the uncertainty bands close to the prior: the
observation has moved the location of the predictive distribution more than its
spread. NPE-PFN produces markedly narrower central 90\% intervals, most visibly
for the momentum components $u$ and $v$, while its medians track the LES profiles
closely. The contrast indicates a genuine information gain in output space rather
than a mere re-centring of prior uncertainty. Additional forcings and horizons
are shown in Appendix~\ref{app:additional_ppc} and behave similarly.

\begin{SCfigure}[1.0][h]
    \centering
    \includegraphics[width=0.65\linewidth]{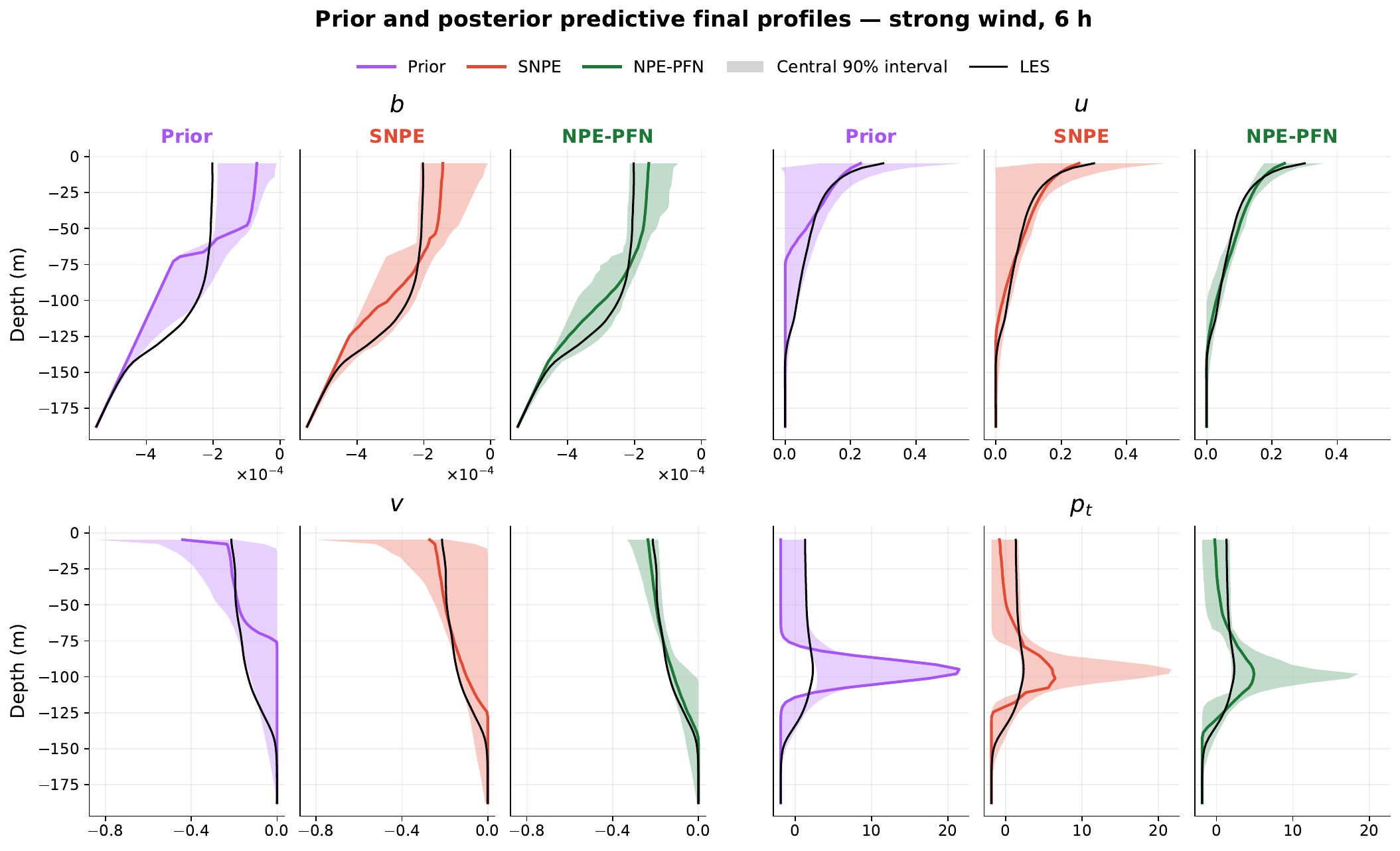}
    \caption{
Prior and posterior predictive final profiles. Colored curves and shaded regions show the predictive medians and central 90\% intervals, while black curves denote the LES profiles. SNPE adjusts the predictive medians but retains uncertainty bands broadly similar to the prior. In contrast, NPE-PFN produces substantially narrower intervals, particularly for $u$ and $v$, indicating a stronger information gain in output space and close tracking of LES profiles.
}
    \label{fig:ppc_strong_wind_6h}
    \label{fig:sidecap}
\end{SCfigure}

\begin{figure}[h]
    \centering
    \includegraphics[width=0.85\linewidth]{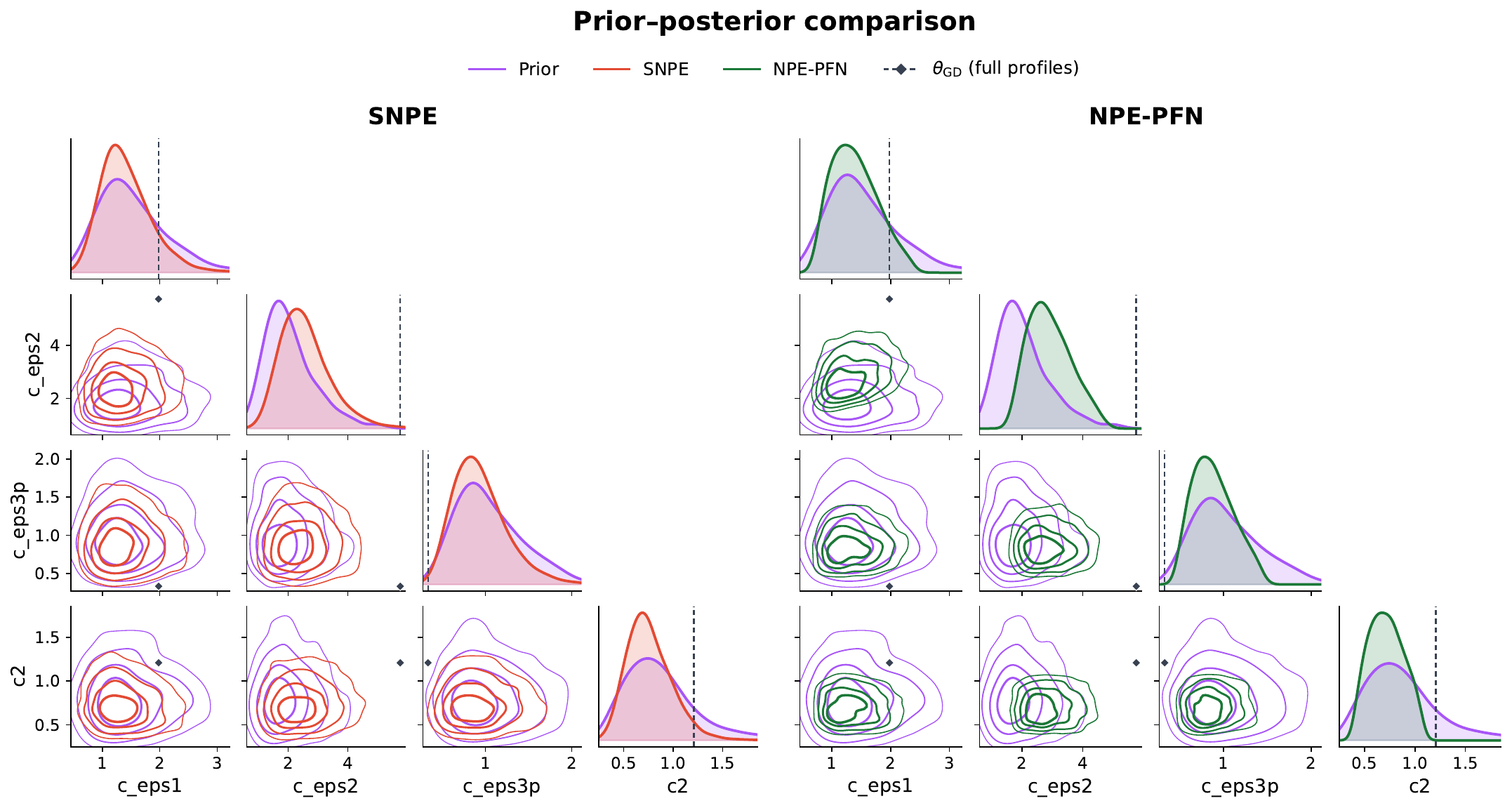}
    \caption{
Prior--posterior comparison in physical parameter space for
$c_{\epsilon1}$, $c_{\epsilon2}$, $c_{\epsilon3}^{+}$, and $c_2$, selected
from the strongest aggregate Spearman correlations across methods detailed in Appendix \ref{app:post_corr_analysis}. Diagonal
panels show marginal KDEs and lower panels show bivariate density contours;
dashed lines and diamonds denote $\boldsymbol{\theta}_{\mathrm{GD}}$ from the
full-profile gradient calibration, which generally lies away from the main
posterior concentration. SNPE remains broadly similar to the prior, whereas
NPE-PFN produces clearer shifts and more concentrated densities, indicating a
stronger posterior update.
}
    \label{fig:prior_posterior_snpe_vs_npe_pfn}
\end{figure}

\textbf{Posterior in parameter space.} Figure~\ref{fig:prior_posterior_snpe_vs_npe_pfn} repeats the comparison in physical parameter space for the four
coefficients with the strongest aggregate Spearman correlations across methods
($c_{\epsilon 1}$, $c_{\epsilon 2}$, $c^{+}_{\epsilon 3}$, $c_2$). The picture is
consistent with the predictive checks: SNPE marginals remain broadly
prior-shaped, whereas NPE-PFN shows clearer location shifts and visibly more
concentrated densities, i.e.\ a stronger posterior update. We also overlay $\boldsymbol{\theta}_{\mathrm{GD}}$, which for several coefficients
falls outside the bulk of the posterior mass. Since the gradient calibration also
attains a lower profile RMSE (Figure~\ref{fig:profile_standardized_rmse_vs_budget}), the posteriors
evidently do not concentrate on the best-fitting configuration. Two factors
plausibly contribute to this: the two procedures optimize different objectives --- full
profiles for $\boldsymbol{\theta}_{\mathrm{GD}}$ against the PCA summary
$\mathbf{s}(\mathbf{x})$ for the posteriors --- and the discrepancy covariance
$\boldsymbol{\Gamma}_{\mathrm{noise}}$, estimated from prior predictive
deviations, is broad enough to leave the likelihood relatively flat.
Disentangling these is left to future work.

\section{Conclusion and future work}

We have presented, to our knowledge, the first application of simulation-based
inference to the calibration of a single-column ocean model, and three
observations follow. First, the TabPFN-based posterior dominates the trained
neural density estimators across all three views: lower RMSE at every budget,
tighter posterior predictives, and a more pronounced update away from the prior.
It achieves this without fitting any parameters, which makes it particularly
attractive in the low-budget regime, i.e.\ whenever each simulator call is
expensive. Second, the resulting calibration is good but not optimal in output
space: the gradient-based estimate remains ahead on RMSE, which is unsurprising,
as it exploits simulator gradients that carry far more information per call than
the input--output pairs available to SBI. The value of the posterior approach
lies elsewhere, in quantifying which parameter configurations are consistent with
the data. Third, the PCA summary statistic was essential rather than cosmetic:
reducing the $3888$-dimensional output to $d_s = 159$ blockwise coefficients,
while preserving the forcing--horizon--variable structure, is what made inference
at these budgets possible.

Several directions follow naturally. The most immediate is to exploit the
differentiability of the JAX implementation---precluded by earlier Fortran
codes---through gradient-guided proposals or hybrid objectives, combining the
per-call information of $\boldsymbol{\theta}_{\mathrm{GD}}$ with the uncertainty
quantification of SBI to reduce the simulation budget further. Non-amortized
inference offers a complementary saving, since sequential and focused posteriors
waste fewer simulations, and these efficiency gains compound. Both matter for the
longer-term goal of scaling to other ocean and climate
simulators, where calibration is harder and each forward run is far costlier than
in \texttt{tunax}.

\section*{Acknowledgements}
This work was financed in part by the Coordenação de Aperfeiçoamento de Pessoal de Nível Superior - Brasil (CAPES)
- Finance Code 001. L.M.C.C is grateful for the fellowship
provided by São Paulo Research Foundation (FAPESP),
grant 2025/06168-8. JLS and PLCR is grateful for the financial support by the French National Research Agency (Agence Nationale de la Recherche) attributed to the SBI4C project of the MIAI AI Cluster, under the reference ANR-23-IACL-0006.

\small
\bibliographystyle{apalike}
\bibliography{biblio}

@article{schneider2017earth,
  title = {Earth System Modeling 2.0: A Blueprint for Models That Learn From Observations and Targeted High‐Resolution Simulations},
  volume = {44},
  ISSN = {1944-8007},
  url = {http://dx.doi.org/10.1002/2017GL076101},
  DOI = {10.1002/2017gl076101},
  number = {24},
  journal = {Geophysical Research Letters},
  publisher = {American Geophysical Union (AGU)},
  author = {Schneider,  Tapio and Lan,  Shiwei and Stuart,  Andrew and Teixeira,  João},
  year = {2017},
  month = Dec 
}

@article{kennedy2001bayesian,
  title={Bayesian calibration of computer models},
  author={Kennedy, Marc C and O'Hagan, Anthony},
  journal={Journal of the Royal Statistical Society: Series B (Statistical Methodology)},
  volume={63},
  number={3},
  pages={425--464},
  year={2001},
  publisher={Wiley Online Library}
}

@article{jolliffe2016principal,
  author  = {Jolliffe, Ian T. and Cadima, Jorge},
  title   = {Principal Component Analysis: A Review and Recent Developments},
  journal = {Philosophical Transactions of the Royal Society A:
             Mathematical, Physical and Engineering Sciences},
  year    = {2016},
  volume  = {374},
  number  = {2065},
  pages   = {20150202},
  doi     = {10.1098/rsta.2015.0202}
}

@article{reffray2015modelling,
  title={Modelling turbulent vertical mixing sensitivity using a 1-D version of NEMO},
  author={Reffray, Guillaume and Bourdalle-Badie, Romain and Calone, Christophe},
  journal={Geoscientific Model Development},
  volume={8},
  number={1},
  pages={69--86},
  year={2015},
  publisher={Copernicus GmbH G{\"o}ttingen, Germany}
}

@article{aldebert2021fast,
  title={A fast and generic method to identify parameters in complex and embedded geophysical models: The example of turbulent mixing in the ocean},
  author={Aldebert, Clement and Koenig, Guillaume and Baklouti, Melika and Frauni{\'e}, Philippe and Devenon, Jean-Luc},
  journal={Journal of Advances in Modeling Earth Systems},
  volume={13},
  number={8},
  pages={e2020MS002245},
  year={2021},
  publisher={Wiley Online Library}
}

@article{wagner2025formulation,
  title={Formulation and calibration of CATKE, a one-equation parameterization for microscale ocean mixing},
  author={Wagner, Gregory LeClaire and Hillier, Adeline and Constantinou, Navid C and Silvestri, Simone and Souza, Andre and Burns, Keaton J and Hill, Chris and Campin, Jean-Michel and Marshall, John and Ferrari, Raffaele},
  journal={Journal of Advances in Modeling Earth Systems},
  volume={17},
  number={4},
  pages={e2024MS004522},
  year={2025},
  publisher={Wiley Online Library}
}

@article{papamakarios2021nf,
  author  = {George Papamakarios and Eric Nalisnick and Danilo Jimenez Rezende and Shakir Mohamed and Balaji Lakshminarayanan},
  title   = {Normalizing Flows for Probabilistic Modeling and Inference},
  journal = {Journal of Machine Learning Research},
  year    = {2021},
  volume  = {22},
  number  = {57},
  pages   = {1--64},
  url     = {http://jmlr.org/papers/v22/19-1028.html}
}

@misc{wagner2026reference,
  author       = {Wagner, G. L.},
  title        = {Reference Large-Eddy Simulations of the upper-ocean boundary layer from the CATKE calibration suite (Wagner et al., 2025)},
  year         = {2026},
  howpublished = {Zenodo},
  note         = {Version 1.0.0. DOI: 10.5281/zenodo.20057136},
  url          = {https://doi.org/10.5281/zenodo.20057136}
}

@inproceedings{deistler2022truncated,
  title={Truncated proposals for scalable and hassle-free simulation-based inference},
  author={Deistler, Michael and Gon{\c{c}}alves, Pedro J. and Macke, Jakob H.},
  booktitle={Advances in Neural Information Processing Systems},
  volume={35},
  pages={23135--23149},
  year={2022}
}

@article{gjini2025ensemble,
  title={The ensemble kalman inversion race},
  author={Gjini, Rebecca and Morzfeld, Matthias and Dunbar, Oliver RA and Schneider, Tapio},
  journal={arXiv preprint arXiv:2511.15853},
  year={2025}
}

@article{blum2013comparative,
  title={A comparative review of dimension reduction methods in approximate Bayesian computation},
  author={Blum, Michael GB and Nunes, Maria Antonieta and Prangle, Dennis and Sisson, Scott A},
  journal={Statistical Science},
  pages={189--208},
  year={2013},
  publisher={JSTOR}
}

@misc{tunaxurl,
  author       = {Mouttapa, Gabriel and Le Sommer, Julien},
  title        = {\texttt{tunax}: Automatic calibration of vertical ocean physics in JAX},
  year         = {2026},
  howpublished = {Zenodo},
  note         = {Version v0.2.11. DOI: 10.5281/zenodo.22143348},
  url          = {https://doi.org/10.5281/zenodo.22143348}
}

@article{reichl2024improving,
  title={Improving equatorial upper ocean vertical mixing in the NOAA/GFDL OM4 model},
  author={Reichl, Brandon G and Wittenberg, Andrew T and Griffies, Stephen M and Adcroft, Alistair},
  journal={Earth and Space Science},
  volume={11},
  number={10},
  pages={e2023EA003485},
  year={2024},
  publisher={Wiley Online Library}
}

@article{hartung2018ec,
  title={An EC-Earth coupled atmosphere--ocean single-column model (AOSCM. v1\_EC-Earth3) for studying coupled marine and polar processes},
  author={Hartung, Kerstin and Svensson, Gunilla and Struthers, Hamish and Deppenmeier, Anna-Lena and Hazeleger, Wilco},
  journal={Geoscientific Model Development},
  volume={11},
  number={10},
  pages={4117--4137},
  year={2018},
  publisher={Copernicus Publications G{\"o}ttingen, Germany}
}

@article{wood2010statistical,
  title={Statistical inference for noisy nonlinear ecological dynamic systems},
  author={Wood, Simon N},
  journal={Nature},
  volume={466},
  number={7310},
  pages={1102--1104},
  year={2010},
  publisher={Nature Publishing Group UK London}
}

@article{cranmer2020frontier,
  title={The frontier of simulation-based inference},
  author={Cranmer, Kyle and Brehmer, Johann and Louppe, Gilles},
  journal={Proceedings of the National Academy of Sciences},
  volume={117}, number={48}, pages={30055--30062}, year={2020}
}

@article{Calvello_Reich_Stuart_2025, title={Ensemble Kalman methods: A mean-field perspective}, volume={34}, DOI={10.1017/S0962492924000060}, journal={Acta Numerica}, author={Calvello, Edoardo and Reich, Sebastian and Stuart, Andrew M.}, year={2025}, pages={123–291}}

@article{Iglesias_2013,
   title={Ensemble Kalman methods for inverse problems},
   volume={29},
   ISSN={1361-6420},
   url={http://dx.doi.org/10.1088/0266-5611/29/4/045001},
   DOI={10.1088/0266-5611/29/4/045001},
   number={4},
   journal={Inverse Problems},
   publisher={IOP Publishing},
   author={Iglesias, Marco A and Law, Kody J H and Stuart, Andrew M},
   year={2013},
   month=Mar, pages={045001} }

@inproceedings{
vetter2025effortless,
title={Effortless, Simulation-Efficient Bayesian Inference using Tabular Foundation Models},
author={Julius Vetter and Manuel Gloeckler and Daniel Gedon and Jakob H. Macke},
booktitle={The Thirty-ninth Annual Conference on Neural Information Processing Systems},
year={2025},
url={https://openreview.net/forum?id=kN0YHWGDPH}
}

@article{deistler2025simulation,
  title={Simulation-based inference: A practical guide},
  author={Deistler, Michael and Boelts, Jan and Steinbach, Peter and Moss, Guy and Moreau, Thomas and Gloeckler, Manuel and Rodrigues, Pedro LC and Linhart, Julia and Lappalainen, Janne K and Miller, Benjamin Kurt and others},
  journal={arXiv preprint arXiv:2508.12939},
  year={2025}
}

@inproceedings{greenberg2019automatic,
  title={Automatic posterior transformation for likelihood-free inference},
  author={Greenberg, David and Nonnenmacher, Marcel and Macke, Jakob},
  booktitle={International conference on machine learning},
  pages={2404--2414},
  year={2019},
  organization={PMLR}
}

@article{Umlauf2005,
  title = {Second-Order Turbulence Closure Models for Geophysical Boundary Layers. {{A}} Review of Recent Work},
  author = {Umlauf, Lars and Burchard, Hans},
  year = 2005,
  month = may,
  journal = {Continental Shelf Research},
  series = {Recent {{Developments}} in {{Physical Oceanographic Modelling}}: {{Part II}}},
  volume = {25},
  number = {7},
  pages = {795--827},
  issn = {0278-4343},
  doi = {10.1016/j.csr.2004.08.004},
  urldate = {2025-05-23}
}


\newpage
\appendix

\section{Prior, Simulator and summary statistics}
\label{app:prior_simulator_summary_stats}
This appendix provides additional details on the prior specification and its prior predictive checks (Section~\ref{app:prior_setup}), the construction of the blockwise PCA summary statistic from prior predictive simulations (Section~\ref{app:pca}), and the Gaussian noise model adopted for the simulator outputs during calibration (Section~\ref{app:noise}).
\subsection{Prior and Simulator Details}
\label{app:prior_setup}

The $k$--$\varepsilon$ parameterization \citep{Umlauf2005} contains
$d_\theta=17$ parameters. It is a two-equation closure that solves one
differential equation for the turbulent kinetic energy (TKE), $k$, and
another for its dissipation rate, $\varepsilon$. The parameters are grouped
according to their role in the closure:
\begin{itemize}
    \item $c_1$, $c_2$, $c_3$, $c_4$, and $c_6$ control the dissipation of
    the pressure--velocity correlation tensor;

    \item $c_{b1}$, $c_{b2}$, $c_{b3}$, $c_{b5}$, and $c_{bb}$ control the
    dissipation of the buoyancy--velocity correlation tensor;

    \item $\sigma_k$ and $\sigma_\varepsilon$ are strictly positive Schmidt
    numbers;

    \item $c_\mu^0$ relates the mixing length to the dissipation rate;

    \item $c_{\varepsilon1}$, $c_{\varepsilon2}$,
    $c_{\varepsilon3}^{-}$, and $c_{\varepsilon3}^{+}$ modify the
    $\varepsilon$ equation.
\end{itemize}

Except for $c_{\varepsilon3}^{-}$, these parameters are constrained to be
positive, allowing their prior distributions to be represented as
log-normal. For each positive parameter, we define
\begin{equation}
    \theta_i \sim
    \operatorname{LogNormal}\!\left(
        \log\theta_{\mathrm{default},i},\,\log(1.5)^2
    \right),
    \qquad \theta_i>0.
\end{equation}
For $c_{\varepsilon3}^{-}$, which is constrained to be negative, the same
prior is placed on its magnitude:
\begin{equation}
    -c_{\varepsilon3}^{-} \sim
    \operatorname{LogNormal}\!\left(
        \log|c_{\varepsilon3,\mathrm{default}}^{-}|,\,
        \log(1.5)^2
    \right).
\end{equation}
Thus, its physical-space prior is log-normal up to the fixed negative sign. Here, $\theta_{\mathrm{default}}$ denotes the customary
\texttt{tunax} parameter configuration listed in
Table~\ref{tab:tunax_default_parameters}. By construction, these values are the prior medians in physical parameter space.
\begin{table}[h]
    \centering
    \caption{Customary \texttt{tunax} parameter values defining the
    physical-space prior medians.}
    \label{tab:tunax_default_parameters}
    \small
    \begin{tabular}{lc@{\qquad}lc}
        \toprule
        Parameter & Default value & Parameter & Default value \\
        \midrule
        $c_1$                    & $5.0000$  &
        $c_{b1}$                 & $5.9500$  \\
        $c_2$                    & $0.8000$  &
        $c_{b2}$                 & $0.6000$  \\
        $c_3$                    & $1.9680$  &
        $c_{b3}$                 & $1.0000$  \\
        $c_4$                    & $1.1360$  &
        $c_{b5}$                 & $0.3333$  \\
        $c_6$                    & $0.4000$  &
        $c_{bb}$                 & $0.7200$  \\
        $c_\mu^0$                & $0.5477$  &
        $\sigma_k$               & $1.0000$  \\
        $\sigma_\varepsilon$     & $1.3000$  &
        $c_{\varepsilon1}$       & $1.4400$  \\
        $c_{\varepsilon2}$       & $1.9200$  &
        $c_{\varepsilon3}^{-}$   & $-0.4000$ \\
        $c_{\varepsilon3}^{+}$   & $1.0000$  &
                                 &           \\
        \bottomrule
    \end{tabular}
\end{table}

For computation, this prior is represented as a diagonal Gaussian in the
unconstrained coordinates
\begin{equation}
    z_i=
    \begin{cases}
        \log(\theta_i), & i\neq i_{-},\\
        \log(-\theta_i), & i=i_{-},
    \end{cases}
\end{equation}
where $i_{-}$ denotes the index of $c_{\varepsilon3}^{-}$, the only
necessarily negative parameter. The transformed prior is
\begin{equation}
    \boldsymbol z \sim
    \mathcal N(\boldsymbol\mu_0,\boldsymbol\Sigma_0),
    \qquad
    \mu_{0,i}=\log|\theta_{\mathrm{default},i}|,
    \qquad
    \boldsymbol\Sigma_0=\log(1.5)^2\boldsymbol I_{17}.
\end{equation}

In physical parameter space, the adopted log-scale standard deviation implies
that the central 68\% prior interval for the absolute value of each parameter
is
\begin{equation}
    \left[
        \frac{|\theta_{\mathrm{default},i}|}{1.5},
        1.5|\theta_{\mathrm{default},i}|
    \right].
\end{equation}
This choice balances informativeness and exploration: it concentrates prior
probability around an established closure configuration while retaining
substantial probability for departures from it.

\begin{figure}[h]
    \centering
    \includegraphics[width=1\linewidth]{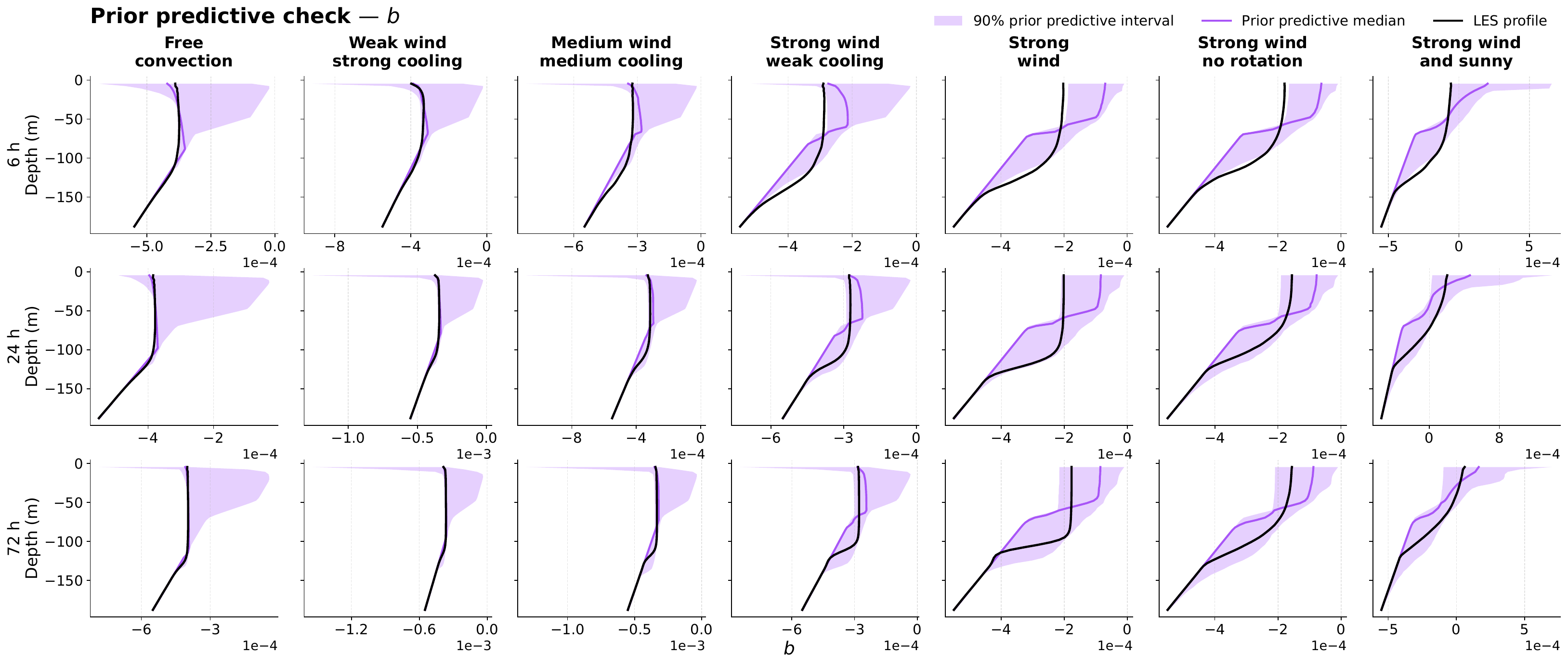}
    \caption{Prior predictive final profiles for buoyancy $b$.  Columns show
    forcing regimes and rows show 6, 24, and 72 hours. At 24 and 72 hours, the central 90\% prior predictive interval generally encompasses the LES profiles. At 6 hours, mild departures occur over some depths and forcing
    regimes.}
    \label{fig:prior_predictive_b}
\end{figure}

\begin{figure}[h]
    \centering
    \includegraphics[width=1\linewidth]{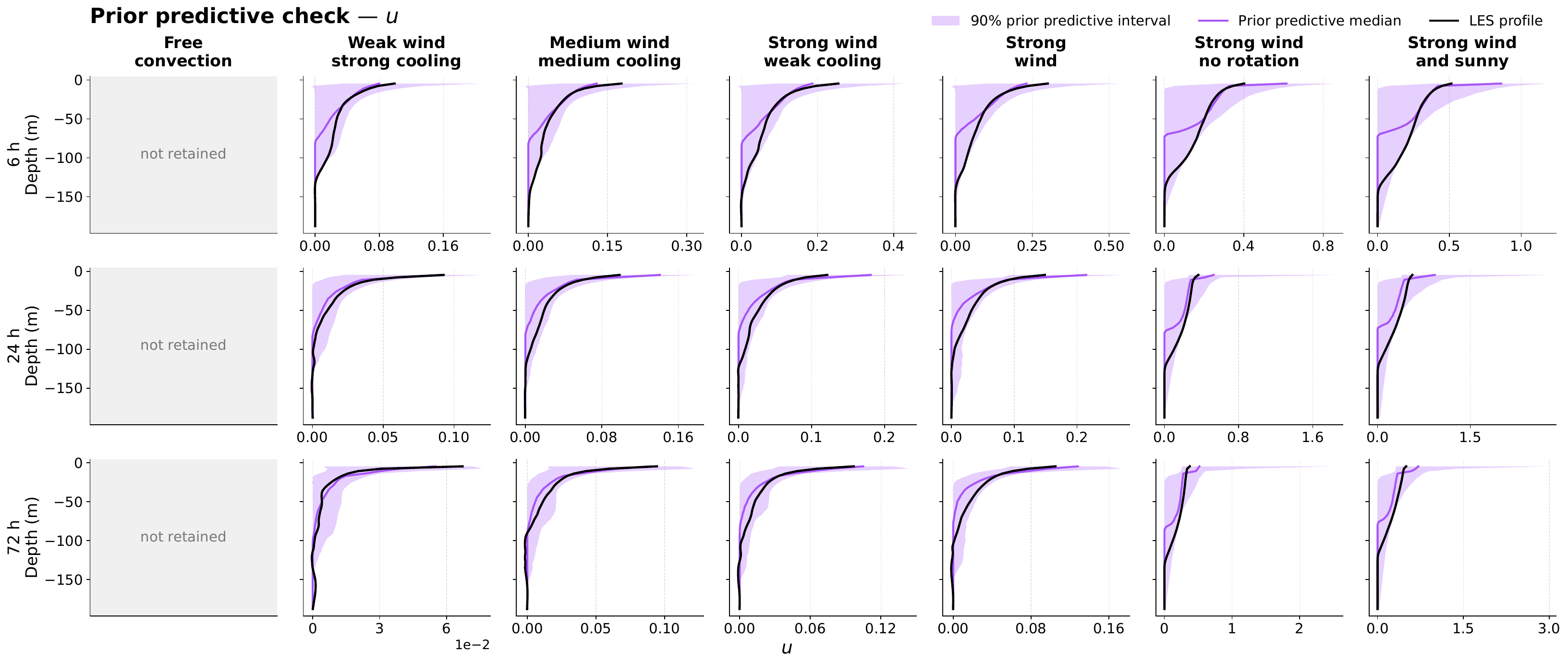}
    \caption{Prior predictive final profiles for the along-wind velocity $u$. The central 90\% prior predictive interval encompasses the LES profiles over most retained forcing and depth combinations, with localized departures in some cases.  Free convection is not retained because it has no imposed wind and therefore no informative $u$ response.}
    \label{fig:prior_predictive_u}
\end{figure}

\begin{figure}[h]
    \centering
    \includegraphics[width=1\linewidth]{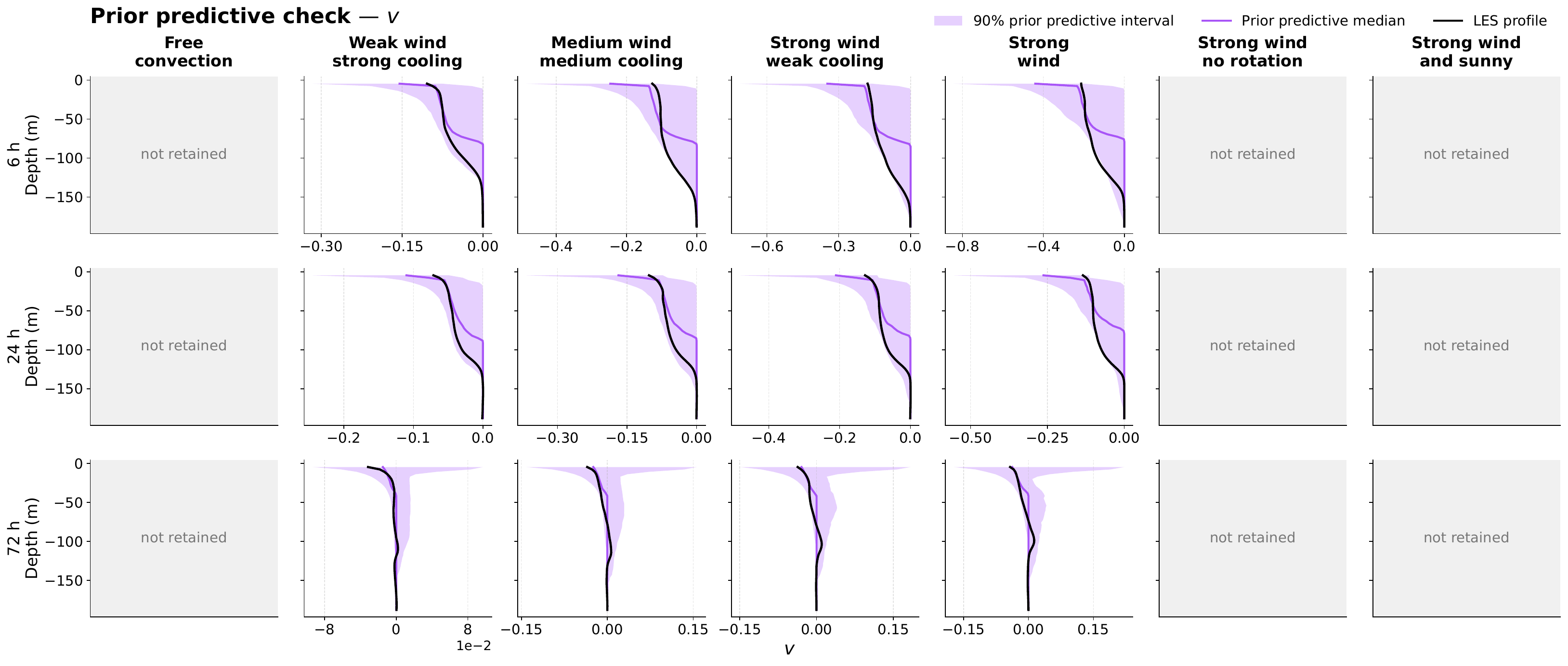}
    \caption{Prior predictive final profiles for the transverse velocity $v$.
    Across the retained rotating and wind-forced configurations, the central
    90\% prior predictive interval encompasses most of the LES profiles.  Free
    convection and the two zero-Coriolis configurations are not retained
    because their transverse profiles provide negligible information for
    calibration.}
    \label{fig:prior_predictive_v}
\end{figure}

\begin{figure}[h]
    \centering
    \includegraphics[width=1\linewidth]{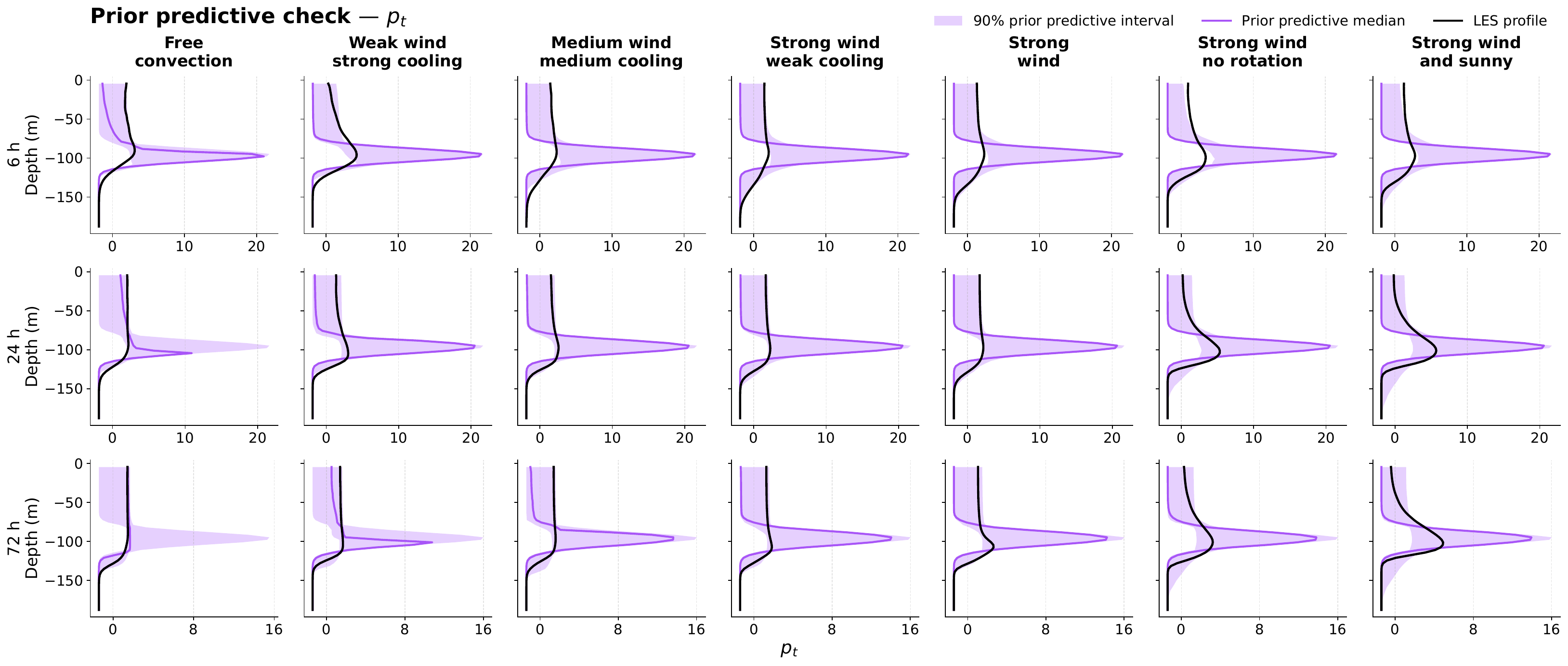}
    \caption{Prior predictive final profiles for the passive tracer $p_t$.
    The central 90\% prior predictive interval generally encompasses a larger
    portion of the LES profiles at 24 and 72 hours.  More visible departures
    occur during the 6-hour transient, showing that the prior predictive
    coverage of the tracer varies throughout the simulated evolution.}
    \label{fig:prior_predictive_pt}
\end{figure}

To assess this trade-off in output space, we construct prior predictive checks
from $M=1000$ independent parameter draws evaluated with the deterministic
\texttt{tunax} simulator.  Figures~\ref{fig:prior_predictive_b}--\ref{fig:prior_predictive_pt} show the final vertical profiles at 6, 24, and
72 hours for seven forcing regimes and 54 calibration depths.  At each depth,
the gray region is the central 90\% prior predictive interval, the gray curve
is its median, and the black curve is the corresponding large-eddy-simulation
(LES) profile. The generally broad bands cover the LES profiles over most
retained depths and forcing regimes, showing that the prior supports a diverse
range of simulator responses.  Nevertheless, they retain clear depth- and
forcing-dependent structure rather than becoming uniformly diffuse, indicating
that the prior remains informative. 

Only physically informative forcing--variable combinations enter inference.
Under free convection there is no imposed wind, so the $u$ and $v$ profiles
have negligible informative variation and are removed.  The transverse
velocity $v$ is also removed for \texttt{strong\_wind\_no\_rotation} and
\texttt{strong\_wind\_and\_sunny}: both configurations have zero Coriolis
parameter and therefore do not generate an informative transverse response.
These exclusions form an explicit, pre-specified simulator mask rather than a post-inference selection. After masking, 72
$(\text{time},\text{forcing},\text{variable})$ blocks remain, giving an output dimension of $72\times54=3888$.

Overall, the prior predictive bands encompass a substantial range of vertical profile behaviour and cover the LES profiles across most retained forcing and depth combinations.  Coverage is less complete for $b$ and $p_t$ at 6 hours, where departures occur over parts of the water column.  The momentum profiles, as well as $b$ and $p_t$ at later times, are more consistently encompassed by the predictive bands.  The checks therefore indicate that the prior provides meaningful exploratory variability without obscuring visible model--data discrepancies.

\subsection{PCA summary statistic---details}
\label{app:pca}

After masking the uninformative forcing--variable combinations, flattening the
72 profiles with 54 depth levels produces an output vector with $D=3888$.
This representation is high-dimensional and redundant because depths within
each profile vary dependently. PCA replaces these correlated coordinates with
a smaller set of coefficients that retains the dominant prior predictive
variation \citep{jolliffe2016principal}, reducing the dimension presented to
the inference methods. We fit PCA separately for each
time-horizon--forcing--variable block, preserving the profile structure and
preventing differences in scale between variables from determining a single
global basis.

We index the retained blocks by $i=1,\ldots,72$. For block $i$, let
$\mathbf{y}^{(m)}_i\in\mathbb{R}^{54}$ denote the profile generated by prior
predictive simulation $m=1,\ldots,M$, with $M=1000$. After subtracting the
blockwise mean $\boldsymbol{\mu}_i$, the centered profiles form a matrix
$\mathbf{X}_i\in\mathbb{R}^{M\times54}$. We compute the singular value
decomposition
\begin{equation}
    \mathbf{X}_i
    =
    \mathbf{U}_i
    \mathbf{D}_i
    \mathbf{V}_i^\top,
\end{equation}
where $d_{i,j}$ is the $j$-th singular value and
$\boldsymbol{\phi}_{i,j}$, the $j$-th column of $\mathbf{V}_i$, is the
corresponding PCA basis vector across depth. The profiles are centered but not
standardized separately at each depth, so the components capture directions
of largest prior predictive variance within each block.

The number of retained components is the smallest value for which the
cumulative explained variance reaches 90\%:
\begin{equation}
    R_i(k)
    =
    \frac{\sum_{j=1}^{k}d_{i,j}^{2}}
         {\sum_{j=1}^{54}d_{i,j}^{2}},
    \qquad
    k_i
    =
    \min\left\{
        k\in\{1,\ldots,5\}:R_i(k)\geq0.9
    \right\}.
\end{equation}
In the present application, every block reaches this threshold using at most
four components, so the upper bound of five is never active.

For a new profile, its $j$-th PCA coefficient is
\begin{equation}
    s_{i,j}(\mathbf{y}_i)
    =
    \boldsymbol{\phi}_{i,j}^{\top}
    \left(\mathbf{y}_i-\boldsymbol{\mu}_i\right),
    \qquad j=1,\ldots,k_i.
\end{equation}
The retained coefficients are stacked following a fixed ordering of the 72
blocks. This reduces the simulator output from 3888 depth-resolved values to
159 PCA coefficients.

\begin{figure}[h]
    \centering
    \includegraphics[width=0.8\linewidth]{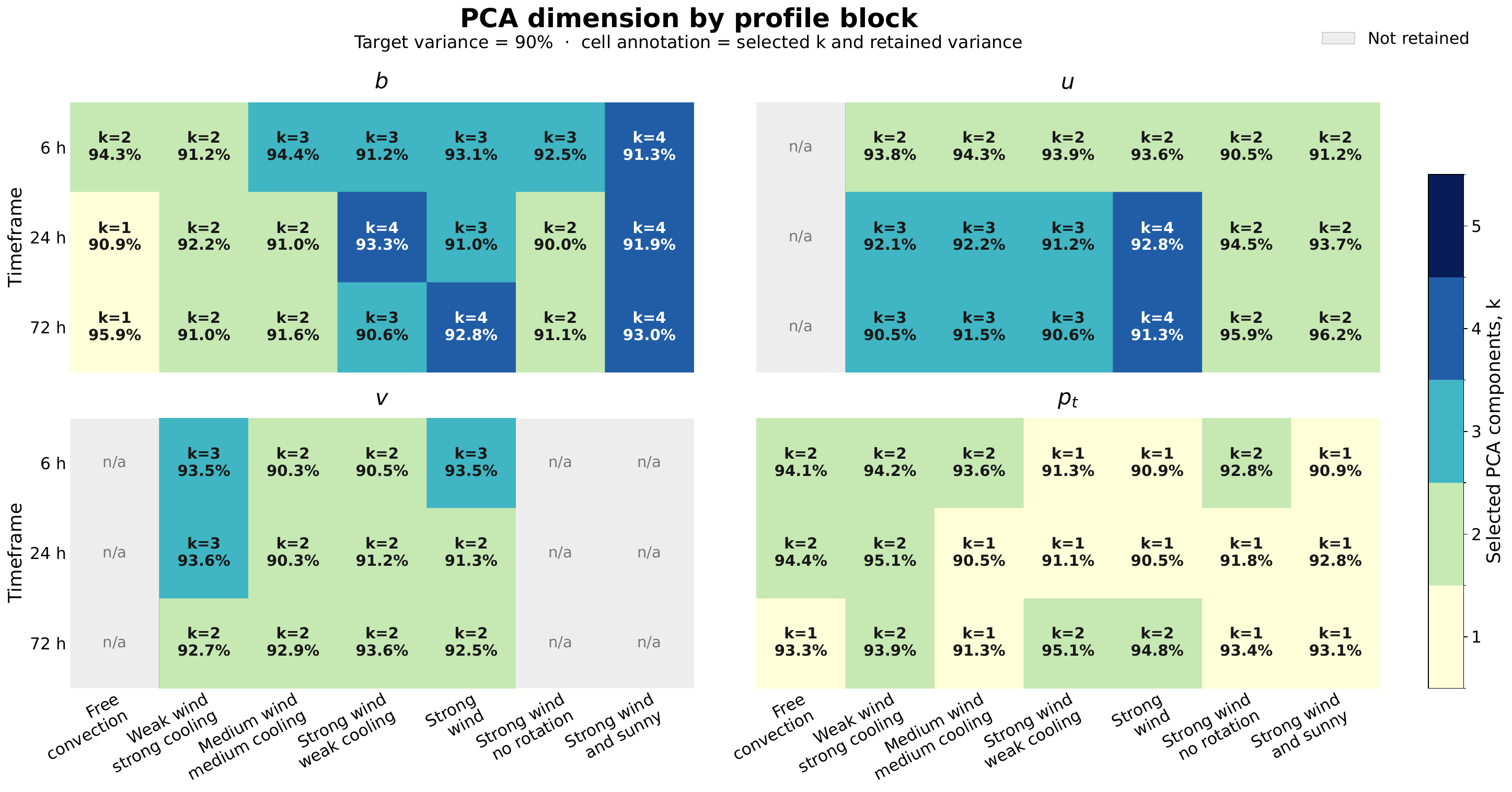}
    \caption{
Blockwise PCA dimensions selected using a 90\% explained-variance threshold.
Panels denote physical variables, rows denote time horizons, and columns
denote forcing regimes; cells report the selected dimension and retained
variance, while gray cells indicate masked combinations. Overall, $b$
generally requires a larger PCA basis than $p_t$, supporting an adaptive
block-specific truncation.
}
    \label{fig:pca_components_by_block}
\end{figure}

Figure~\ref{fig:pca_components_by_block} shows that the selected dimension
depends on both the physical variable and the forcing--horizon combination.
Most buoyancy blocks require two to four components. The velocity profiles
are moderately compressible: $u$ generally requires two or three components
and occasionally reaches four, whereas $v$ is mostly represented by two
components, with three required in some blocks. In contrast, one or two
components are generally sufficient for $p_t$. This variation motivates
selecting $k$ separately for each block, retaining additional PCA directions
when the profile variability is less concentrated without introducing
unnecessary summary coefficients for more compressible blocks.

\subsection{Simulator noise model}
\label{app:noise}

Conditional on its closure parameters, \texttt{tunax} defines a deterministic
mapping from parameters to vertical profiles. Nevertheless, its output should
not be expected to reproduce the higher-fidelity LES exactly. Differences may
arise from structural approximations in the turbulence closure, uncertainty
induced by the adopted parameterization, unresolved or numerical effects, and
uncertainty associated with the reference data. Rather than modeling these
contributions separately, we represent their combined effect through a single
effective Gaussian discrepancy term, following standard observation models
used in inverse problems and simulator calibration
\citep{Iglesias_2013,kennedy2001bayesian,wood2010statistical}.

The discrepancy is introduced after the blockwise PCA transformation. Let
$F(\boldsymbol\theta)$ denote the retained \texttt{tunax} profiles and
$\mathbf{s}$ the fitted PCA summary operator. We assume
\begin{equation}
    \mathbf{s}_{\mathrm{obs}}
    =
    \mathbf{s}\!\left(F(\boldsymbol\theta)\right)
    + \boldsymbol\eta,
    \qquad
    \boldsymbol\eta
    \sim
    \mathcal{N}\!\left(
        \mathbf{0},
        \boldsymbol\Gamma_{\mathrm{noise}}
    \right),
    \qquad
    \mathbf{s}_{\mathrm{obs}}
    =
    \mathbf{s}(\mathbf{x}_{\mathrm{LES}}).
\end{equation}
For SBI training, independent draws from this multivariate Gaussian are added
to the simulated PCA summaries, while the LES summary remains fixed. The same
covariance determines the discrepancy weighting used by the calibration
procedures.

We estimate the covariance from the same $M=1000$ prior predictive
simulations used to fit the PCA. For each prior draw, we compute
\begin{equation}
    \mathbf{d}^{(m)}
    =
    \mathbf{s}\!\left(F(\boldsymbol\theta^{(m)})\right)
    -
    \mathbf{s}_{\mathrm{obs}},
\end{equation}
and define
\begin{equation}
    \boldsymbol\Gamma_{\mathrm{PCA}}
    =
    \frac{1}{M-1}
    \sum_{m=1}^{M}
    \left(\mathbf{d}^{(m)}-\bar{\mathbf d}\right)
    \left(\mathbf{d}^{(m)}-\bar{\mathbf d}\right)^\top.
\end{equation}
Because the discrepancies are centered and $\mathbf{s}_{\mathrm{obs}}$ is
fixed, this is equivalently the covariance of the prior predictive summaries.
We use it as a practical covariance model for the unresolved discrepancy. The
Gaussian discrepancy is assumed to have zero mean, so any persistent bias
between \texttt{tunax} and LES is not modeled separately.

\begin{figure}[h]
    \centering
    \includegraphics[width=0.8\linewidth]{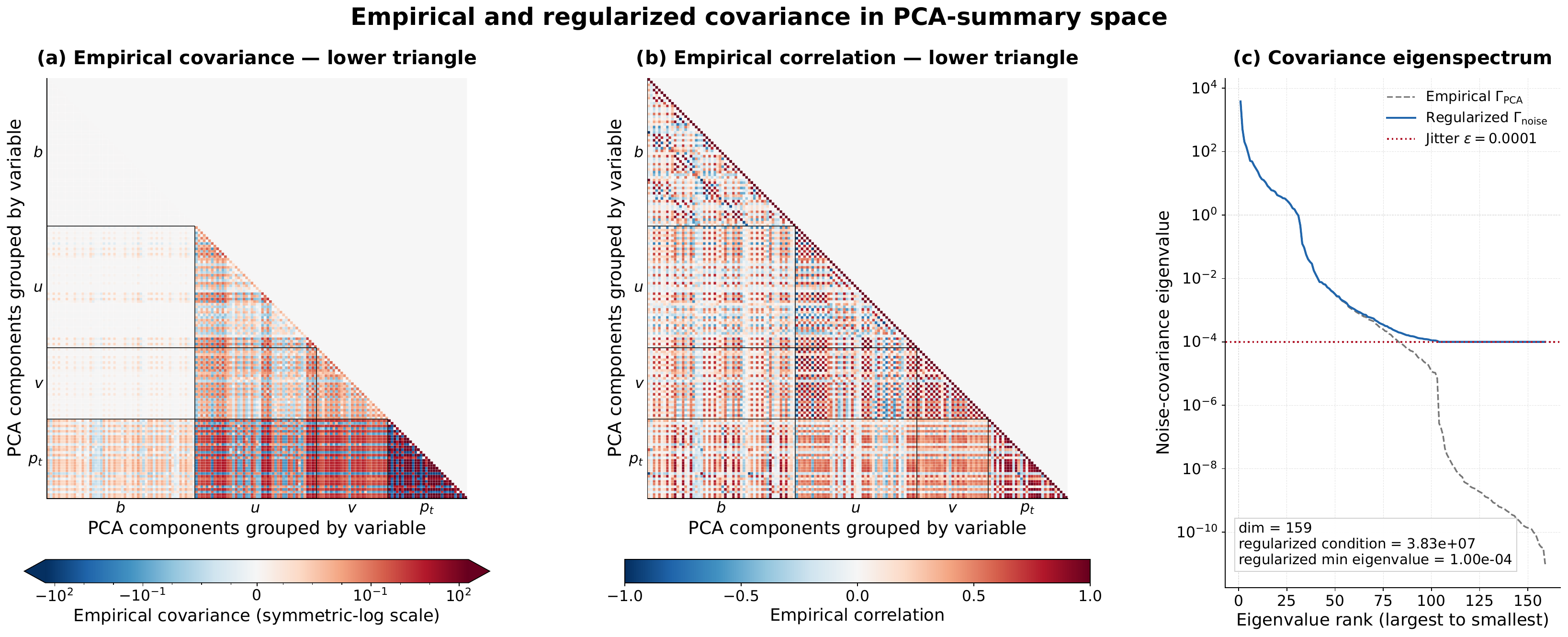}
    \caption{
    Empirical covariance structure in the 159-dimensional PCA-summary space:
    (a) covariance magnitudes, showing nearly zero values for $b$ and the
    largest values for $p_t$; (b) standardized within- and cross-variable
    correlations; and (c) eigenspectra before and after adding the $10^{-4}$
    diagonal nugget. The regularization lifts the low-variance directions
    dominated by $b$ while preserving the dominant covariance structure.
    }
    \label{fig:pca_noise_covariance}
\end{figure}

Panels~(a) and~(b) of Figure~\ref{fig:pca_noise_covariance} provide
complementary views of the empirical covariance. Panel~(a) shows that the
covariance magnitudes associated with $b$ are nearly zero on the global
scale, whereas $u$ and $v$ exhibit intermediate values and $p_t$ dominates
the largest magnitudes. After standardizing these scales, panel~(b) reveals
localized dependencies involving $b$ and more widespread correlations within
and across the other variables. In particular, $p_t$ displays the most
consistently dense within-block structure, with $v$ also showing pronounced
dependence. The cross-block correlations are consistent with common closure
parameters producing coordinated responses across forcing regimes and time
horizons. Thus, the panels describe how the discrepancy is scaled and coupled
during calibration.

Panel~(c) connects this covariance structure to its numerical regularization.
The lower end of the empirical eigenspectrum is dominated by $b$-related
directions with nearly zero variance. For a positive-definite covariance
matrix, the condition number is the ratio between its largest and smallest
eigenvalues,
$\kappa(\boldsymbol\Gamma)=
\lambda_{\max}/\lambda_{\min}$, and measures the numerical sensitivity of its
inversion. A large condition number implies that small numerical or
estimation errors may be amplified and that near-zero-variance directions may
receive excessive weight during calibration. We therefore define
\begin{equation}
    \boldsymbol\Gamma_{\mathrm{noise}}
    =
    \boldsymbol\Gamma_{\mathrm{PCA}}
    +\epsilon_{\mathrm{jit}}\mathbf I,
    \qquad
    \epsilon_{\mathrm{jit}}=10^{-4}.
\end{equation}
This diagonal nugget raises the smallest eigenvalues and stabilizes the
inversion, primarily regularizing the $b$-related directions while leaving
the dominant covariance structure, largely associated with $p_t$,
essentially unchanged.

\section{Additional Results}
\label{app:additional_results}

This appendix complements the main results with further comparisons among the SBI methods. All budget-dependent RMSE analyses consider simulation budgets of $\{50, 100, 200, 400, 600, 800, 1000, 1200, 1800, 2400, 3000\}$. Section~\ref{app:gradient_based_results} compares the SBI posterior means with the gradient-based calibration estimate, both globally through the overall parameter RMSE and individually for each \texttt{tunax} parameter. Section~\ref{app:post_corr_analysis} presents the posterior Spearman correlation matrices for the final SBI methods. Finally, Section~\ref{app:additional_ppc} provides additional posterior predictive comparisons for two further forcing--time combinations.

\subsection{Comparison to gradient-based calibration}
\label{app:gradient_based_results}

We use the full-profile gradient-based calibration
$\boldsymbol\theta_{\mathrm{GD}}$ as a point reference for assessing the
parameter-space location inferred by the SBI methods. For each method and
simulation budget, we compute
\begin{equation}
    \operatorname{RMSE}_{\theta}
    =
    \left[
        \frac{1}{d_\theta}
        \sum_{i=1}^{d_\theta}
        \left(
            \bar{\theta}_i-\theta_{\mathrm{GD},i}
        \right)^2
    \right]^{1/2},
\end{equation}
where $\bar{\boldsymbol\theta}$ is the arithmetic posterior mean in physical
parameter space and $d_\theta=17$. Markers and error bars report the mean and
standard error of this quantity across seeds.

\begin{figure}[h]
    \centering
    \includegraphics[width=0.55\linewidth]
    {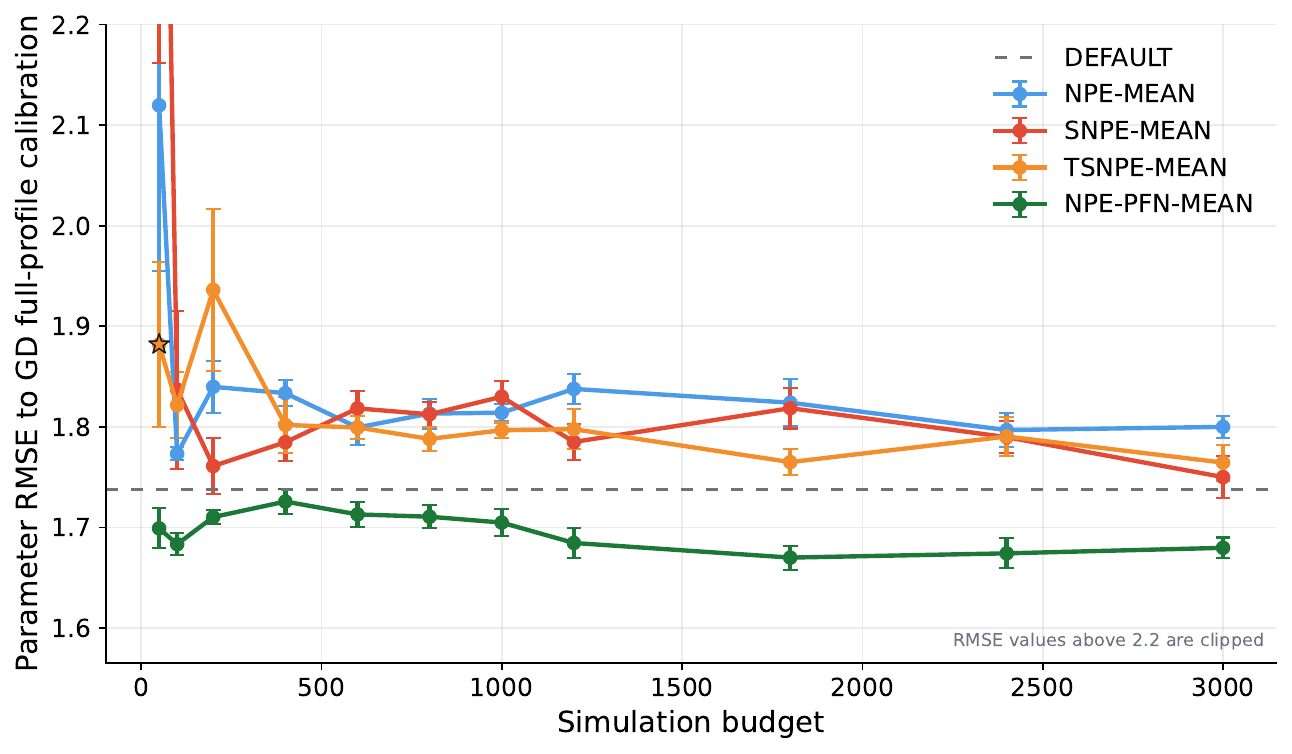}
\caption{
Physical-space RMSE between the posterior mean and
$\boldsymbol\theta_{\mathrm{GD}}$, reported as mean and SEM across seeds.
The dashed line denotes the customary \texttt{tunax} parameters; the starred
TSNPE point at $B=50$ uses a 1\%-trimmed mean, and values above $2.2$ are
clipped. NPE-PFN yields lower RMSE across budgets, while NPE, SNPE, and TSNPE
generally fluctuate slightly above the default-reference distance.
}
    \label{fig:theta_rmse_physical}
\end{figure}

The only exception is TSNPE at $B=50$. At this very small budget, one run
produced an unusually heavy posterior tail, causing a few extreme samples to
dominate the arithmetic mean. For this point, we use a componentwise
1\%-trimmed posterior mean for all seeds, removing the lowest and highest 1\%
of the samples for each parameter. The adjusted point is marked by a star in
Figure~\ref{fig:theta_rmse_physical}. This changes only the displayed point
summary and does not modify the fitted posterior or any density-based
analysis.

For readability, the vertical axis in
Figure~\ref{fig:theta_rmse_physical} is limited to $2.2$, emphasizing the
differences between methods over the moderate- and large-budget regimes.
Values above this limit remain included in all calculations; in particular,
the SNPE result at $B=50$ exceeds the displayed range and is indicated by the
curve entering through the upper boundary.

NPE-PFN remains consistently closer to
$\boldsymbol\theta_{\mathrm{GD}}$ than the flow-based estimators and lies below
the default-reference distance at every budget. Its performance is already
stable at the smallest budgets and exhibits only modest subsequent variation.
NPE, SNPE, and TSNPE show a less systematic dependence on budget and, after
their more variable small-budget behavior, remain clustered near the
default-reference distance. At the largest budgets, SNPE and TSNPE approach
this reference, whereas NPE remains slightly farther away. Increasing the
simulation budget therefore does not necessarily move the posterior mean
monotonically towards $\boldsymbol\theta_{\mathrm{GD}}$, which was obtained
using a distinct full-profile optimization objective detailed in Appendix \ref{app:grad_descente_calib}.

\begin{figure}[h]
    \centering
    \includegraphics[width=0.8\linewidth]
    {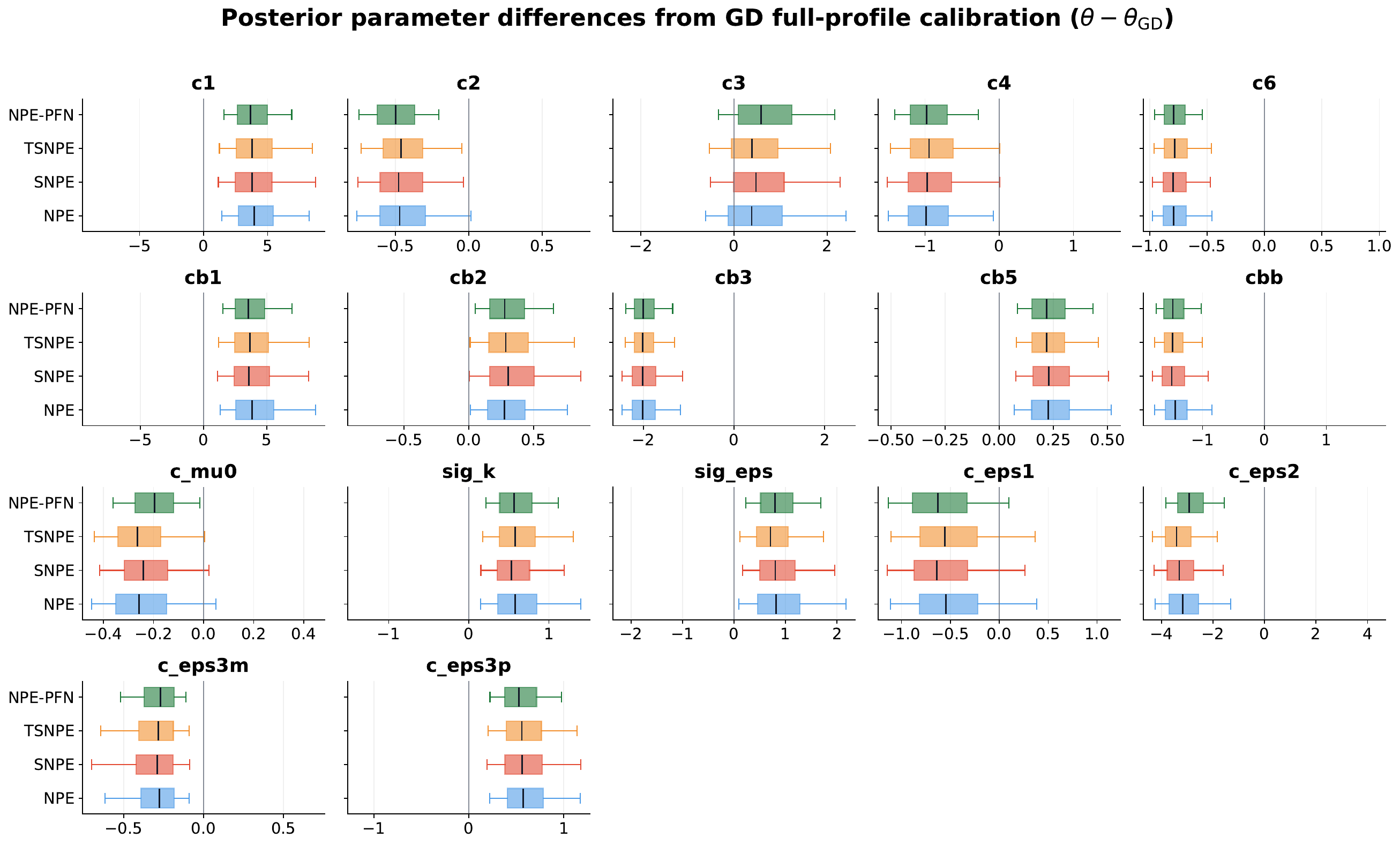}
   \caption{
Posterior parameter differences relative to
$\boldsymbol\theta_{\mathrm{GD}}$ for each method's lowest-profile-RMSE seed
at budget $3000$. NPE, SNPE, and TSNPE largely overlap, whereas NPE-PFN
shows the clearest shifts for $c_{\epsilon2}$ and $c_3$.
}
    \label{fig:parameter_difference_boxplots}
\end{figure}

Because the global RMSE aggregates parameters expressed on different physical
scales, we complement it with the parameter-wise posterior differences in
Figure~\ref{fig:parameter_difference_boxplots}. These boxplots display the
complete sampled posteriors and are unaffected by the adjustment applied to
the TSNPE point at $B=50$.

The parameter-wise comparison shows substantial overlap between NPE, SNPE,
and TSNPE for most coefficients. NPE-PFN differs most clearly for
$c_{\epsilon2}$ and $c_3$, with smaller shifts for $c_{\mu}^{0}$, $c_1$, and
$c_{b1}$. These changes are not uniformly directed towards
$\boldsymbol\theta_{\mathrm{GD}}$: NPE-PFN is closer for coefficients such as
$c_{\mu}^{0}$ and $c_{\epsilon2}$, whereas its $c_3$ distribution is displaced
farther from the gradient reference. Its lower global RMSE therefore results
from the combined parameter-specific changes rather than uniformly closer
agreement for every coefficient.

\subsection{Posterior correlation analysis}
\label{app:post_corr_analysis}

\begin{figure}[h]
    \centering
    \includegraphics[width=0.6\linewidth]
    {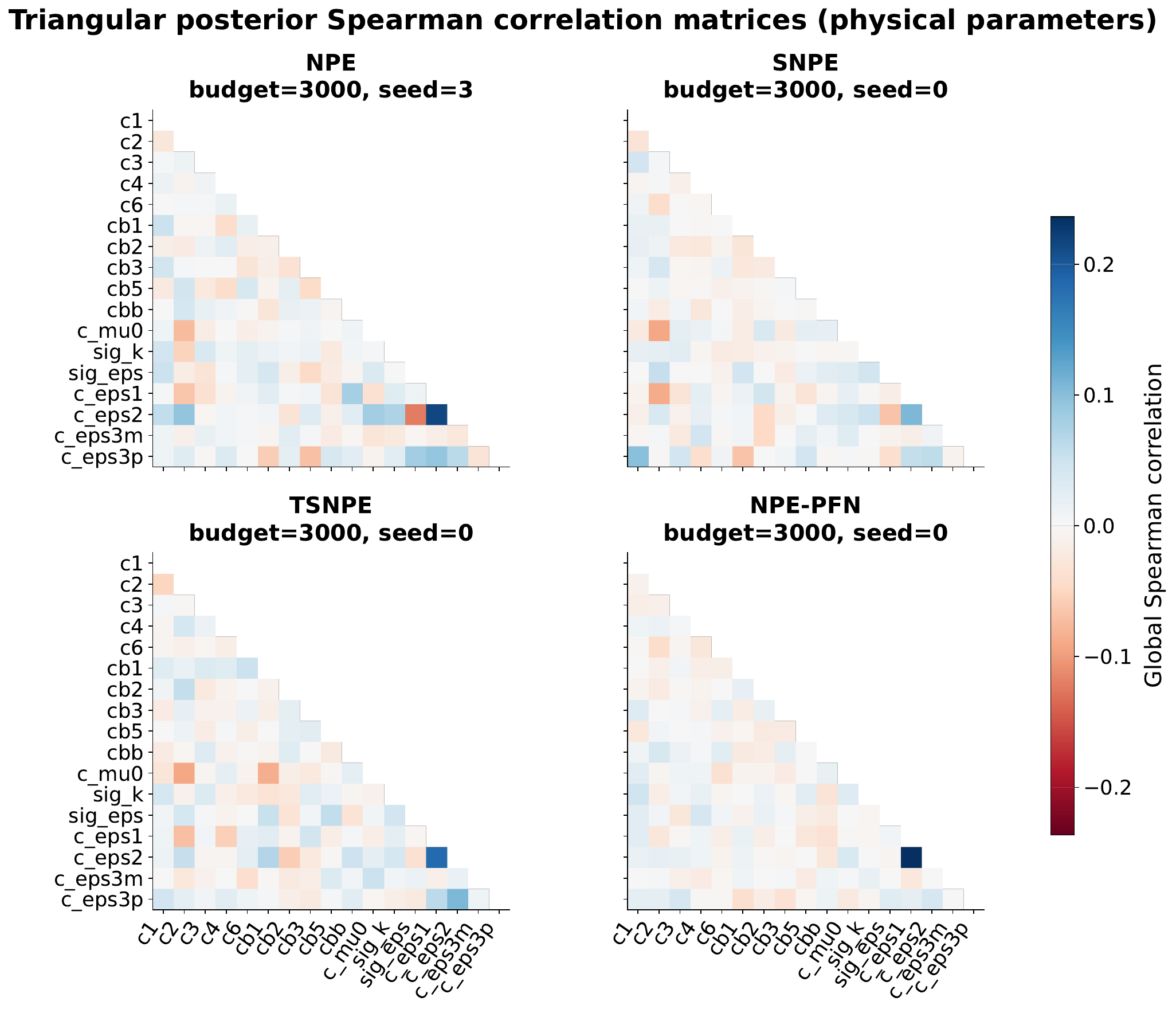}
    \caption{
    Posterior Spearman correlation matrices in physical parameter space for
    the selected runs at a simulation budget of $3000$. All panels share the
    same color scale. Pairwise correlations are generally weak, with the
    positive association between $c_{\varepsilon1}$ and
    $c_{\varepsilon2}$ providing the clearest relationship consistently
    recovered across methods. These matrices motivate the four-parameter
    subset examined in the posterior pairplots.
    }
    \label{fig:posterior_correlations}
\end{figure}

We examine posterior dependence to identify parameter combinations that may
compensate for one another while producing similar simulator responses and to
assess whether such relationships are consistently recovered across inference
methods. We use Spearman correlation because it captures monotonic dependence
without requiring a linear relationship or normally distributed posterior
samples. Figure~\ref{fig:posterior_correlations} shows the resulting
correlation matrices for the selected run of each method at the largest
simulation budget.

Overall, the posterior correlations between parameters are weak, with most
pairs exhibiting Spearman coefficients close to zero across all methods. This
indicates limited pairwise monotonic dependence in the estimated posteriors.
The clearest recurring relationship is the positive association between
$c_{\varepsilon1}$ and $c_{\varepsilon2}$, although its magnitude remains
moderate. Weaker but consistently signed relationships are also observed
between $c_{\varepsilon2}$ and $c_{\varepsilon3}^{+}$, between
$c_{\varepsilon1}$ and $c_{\varepsilon3}^{+}$, and between $c_2$ and the two
former dissipation coefficients. In particular, $c_2$ is weakly negatively
associated with $c_{\varepsilon1}$ and weakly positively associated with
$c_{\varepsilon2}$.

To summarize these dependencies, we rank each parameter pair using its mean absolute Spearman correlation across the $M=4$ inference methods,
\begin{equation}
    R_{ij}
    =
    \frac{1}{M}
    \sum_{m=1}^{M}
    \left|\rho_{S,ij}^{(m)}\right|.
\end{equation}
The method-specific coefficients in
Table~\ref{tab:posterior_correlation_ranking} retain their signs, whereas
$R_{ij}$ measures the average strength of each association without
cancellation across methods. The ranking confirms that
$(c_{\varepsilon1},c_{\varepsilon2})$ is the most prominent common pair and
that the remaining associations are comparatively weak. The four-parameter
set with the largest aggregate within-subset dependence is
$\{c_{\varepsilon1},c_{\varepsilon2},c_{\varepsilon3}^{+},c_2\}$.
We therefore use this subset in the pairplots to examine the corresponding
marginal and joint posterior structures in greater detail.

\begin{table}[ht]
    \centering
    \small
    \caption{
    Parameter pairs with the largest mean absolute posterior Spearman
    correlation across methods. Method-specific coefficients retain their
    signs, while $R_{ij}$ summarizes the average association strength.
    }
    \label{tab:posterior_correlation_ranking}
    \begin{tabular}{lrrrrr}
        \toprule
        Parameter pair & NPE & SNPE & TSNPE & NPE-PFN & $R_{ij}$ \\
        \midrule
        $(c_{\varepsilon1},c_{\varepsilon2})$
            &  0.215 &  0.105 &  0.183 &  0.236 & 0.185 \\
        $(c_{\varepsilon2},c_{\varepsilon3}^{+})$
            &  0.063 &  0.060 &  0.106 &  0.040 & 0.067 \\
        $(c_2,c_{\mu}^{0})$
            & -0.074 & -0.091 & -0.091 & -0.007 & 0.066 \\
        $(c_2,c_{\varepsilon1})$
            & -0.065 & -0.087 & -0.071 & -0.027 & 0.062 \\
        $(c_{\varepsilon1},c_{\varepsilon3}^{+})$
            &  0.092 &  0.057 &  0.062 &  0.021 & 0.058 \\
        $(\sigma_{\varepsilon},c_{\varepsilon2})$
            & -0.120 & -0.068 & -0.035 & -0.006 & 0.057 \\
        $(c_2,c_{\varepsilon2})$
            &  0.093 &  0.035 &  0.055 &  0.020 & 0.051 \\
        $(c_1,c_{\varepsilon3}^{+})$
            &  0.011 &  0.099 &  0.046 &  0.024 & 0.045 \\
        $(c_{\mu}^{0},c_{\varepsilon2})$
            &  0.082 &  0.037 &  0.022 &  0.037 & 0.044 \\
        $(\sigma_{\varepsilon},c_{\varepsilon3}^{+})$
            &  0.082 & -0.040 & -0.022 &  0.028 & 0.043 \\
        \bottomrule
    \end{tabular}
\end{table}

\subsection{Additional Posterior Predictive Checks}
\label{app:additional_ppc}

\begin{figure}[h]
    \centering
    \includegraphics[width=0.85\linewidth]
    {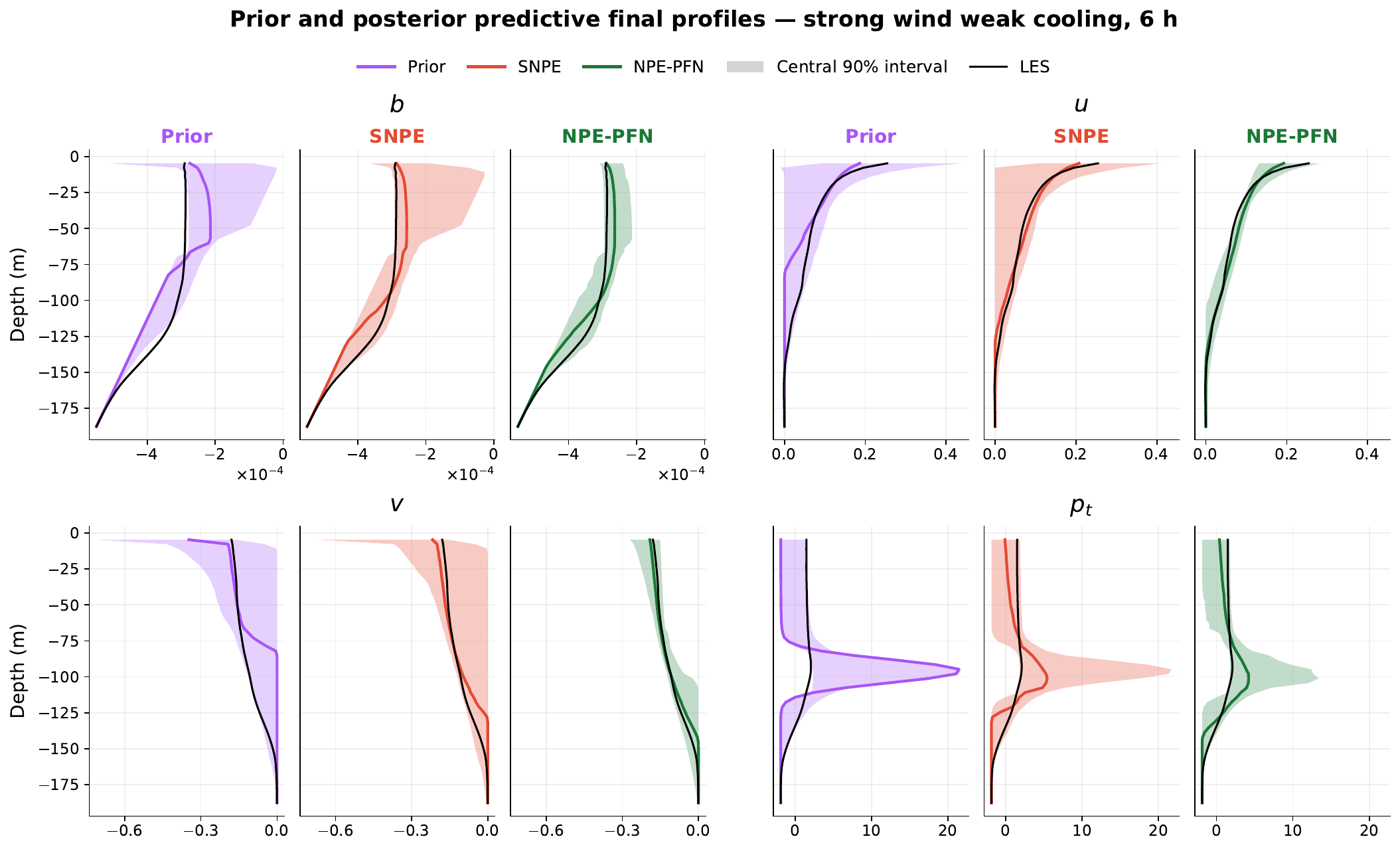}
    \caption{
    Prior and posterior predictive final profiles under strong wind and weak
    cooling at 6 hours. Colored curves and bands denote predictive medians and
    central 90\% intervals, and black curves denote the LES profiles. SNPE
    largely retains the prior predictive spread, whereas NPE-PFN produces
    sharper intervals, particularly for $u$ and $v$, together with clearer
    median adjustments for $b$, $v$, and $p_t$.
    }
    \label{fig:ppc_strong_wind_weak_cooling_6h}
\end{figure}

\begin{figure}[h]
    \centering
    \includegraphics[width=0.85\linewidth]
    {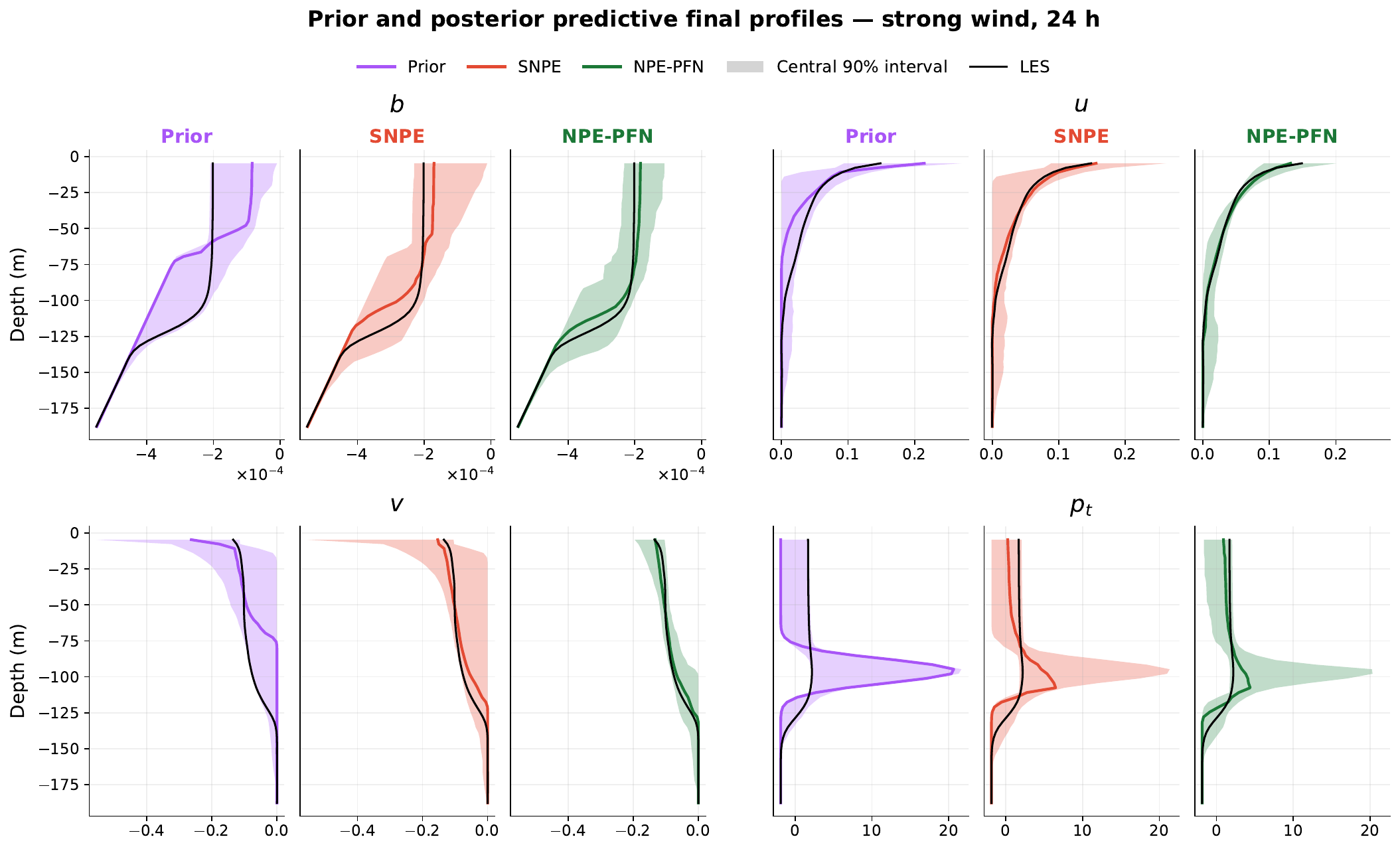}
    \caption{
    Prior and posterior predictive final profiles under strong wind at
    24 hours. NPE-PFN preserves the behavior observed at 6 hours, producing
    the clearest uncertainty contraction for the velocity profiles and
    predictive medians generally closer to the LES reference. SNPE improves
    the median profiles but retains uncertainty bands similar to the prior.
    }
    \label{fig:ppc_strong_wind_24h}
\end{figure}

Figures~\ref{fig:ppc_strong_wind_weak_cooling_6h} and
\ref{fig:ppc_strong_wind_24h} extend the posterior predictive comparison in
Figure 2 to another forcing regime and a later time
horizon. Across all three cases, SNPE shifts the predictive median towards the
LES profiles, but its uncertainty bands remain broadly comparable to those
induced by the prior. NPE-PFN produces a more pronounced posterior update,
with generally sharper predictive intervals and median profiles that more
closely follow the LES reference. This contrast is particularly clear for the
velocity components, indicating that the behavior observed in the main text
is not restricted to a single forcing--time combination.

Under strong wind and weak cooling at 6 hours,
Figure~\ref{fig:ppc_strong_wind_weak_cooling_6h}, the largest NPE-PFN
contraction occurs for $u$ and $v$, with a visible reduction also for $b$.
For $p_t$, the reduction in uncertainty is less pronounced, although the
NPE-PFN median remains closer to the LES profile than the prior median. At
24 hours under strong wind, Figure~\ref{fig:ppc_strong_wind_24h}, the same
ordering persists: NPE-PFN remains sharper than SNPE, especially for $u$ and
$v$, while differences for $b$ and $p_t$ are expressed more through the
location of the predictive median than through a large contraction of the
interval.

\section{Methods Details}
This appendix provides additional implementation details for the SBI methods, including their architectures and hyperparameter configurations
(Section~\ref{app:sbi_details}), as well as the optimization procedure used for the gradient-based calibration (Section~\ref{app:grad_descente_calib}).

\subsection{SBI methods details}
\label{app:sbi_details}

All methods infer the $d_\theta=17$ closure parameters in the unconstrained
coordinates
$\boldsymbol z=T(\boldsymbol\theta)$
defined in Appendix~\ref{app:prior_setup}. Training inputs consist of the
$d_s=159$ blockwise PCA coefficients described in
Appendix~\ref{app:pca}. Independent Gaussian perturbations are added to the
simulated summaries during training according to the observation model in
Appendix~\ref{app:noise}, while inference is conditioned on the fixed LES
summary $\boldsymbol s_{\mathrm{obs}}$.

NPE, SNPE, and TSNPE are implemented with \texttt{sbi} 0.26.1. All three use
a masked autoregressive flow (MAF) to estimate the conditional density
$q_\phi(\boldsymbol z\mid\boldsymbol s)$. The MAF contains five affine
autoregressive transformations, each followed by a random permutation. Every
MADE conditioner has 50 hidden features, two feed-forward blocks, and
$\tanh$ activations. Dropout and batch normalization are disabled. The PCA
coefficients are supplied through an identity embedding, and the parameters
and summaries are standardized coordinatewise using the training data.

The flows are trained using Adam with a learning rate of
$5\times10^{-4}$ and a batch size of 50, for at most 500 epochs. Ten percent
of the simulations are reserved for validation, and training stops if the
validation objective does not improve for 20 epochs. The total gradient norm
is clipped at 5. The standard posterior sample size is $S=10\,000$ draws
conditioned on $\boldsymbol s_{\mathrm{obs}}$.

\paragraph{NPE.}
For the single-round NPE baseline, all $B$ simulations are drawn from the
prior. The estimator is trained by minimizing the conditional negative
log-likelihood
\begin{equation}
    \mathcal{L}_{\mathrm{NPE}}(\phi)
    =
    -\frac{1}{B}
    \sum_{i=1}^{B}
    \log q_\phi
    \left(
        \boldsymbol z^{(i)}
        \mid
        \widetilde{\boldsymbol s}^{(i)}
    \right),
\end{equation}
where $\widetilde{\boldsymbol s}^{(i)}$ denotes the noise-perturbed PCA
summary. This produces an amortized posterior estimator, although it is
evaluated here only at the fixed LES observation.

\paragraph{SNPE.}
SNPE uses the NPE-C/SNPE-C procedure of
\citet{greenberg2019automatic}. The total simulation budget is divided as
evenly as possible among three rounds. The first round draws parameters from
the prior. After each round, the posterior conditioned on
$\boldsymbol s_{\mathrm{obs}}$ becomes the proposal for the following round,
concentrating new \texttt{tunax} simulations in regions with greater
posterior support. After the first round, \texttt{sbi} uses the atomic
SNPE-C objective with its default of 10 atoms to account for the proposal
distribution.

\paragraph{TSNPE.}
TSNPE also uses three approximately equal rounds and the same MAF architecture,
but constructs its proposal by restricting the original prior to a
high-posterior-density region \citep{deistler2022truncated}. After round $r$,
the proposal has the form
\begin{equation}
    \widetilde{p}_r(\boldsymbol z)
    \propto
    p(\boldsymbol z)\,
    \mathbb{I}
    \left[
        q_{\phi_r}
        \left(
            \boldsymbol z
            \mid
            \boldsymbol s_{\mathrm{obs}}
        \right)
        \geq \kappa_r
    \right],
\end{equation}
where $\kappa_r$ is determined from a posterior-density quantile. We use a
quantile of $10^{-4}$ for $B<1200$ and $0.1$ for $B\geq1200$. Samples from the
restricted prior are obtained by rejection sampling. Because this proposal
remains proportional to the original prior within its retained support, the
ordinary NPE maximum-likelihood objective is used in every round.

\paragraph{NPE-PFN.}
NPE-PFN replaces the trained normalizing flow with the pretrained
\texttt{TabPFNRegressor} from TabPFN 8.2.0
\citep{vetter2025effortless}. Its pretrained parameters are not updated using
the \texttt{tunax} simulations. Instead, the simulations form an in-context
regression dataset, so the simulation budget determines the number of
examples supplied to the pretrained model.

The multivariate posterior is represented through the autoregressive
factorization
\begin{equation}
    q
    \left(
        \boldsymbol z
        \mid
        \boldsymbol s_{\mathrm{obs}}
    \right)
    =
    \prod_{j=1}^{d_\theta}
    q_j
    \left(
        z_j
        \mid
        \boldsymbol z_{1:j-1},
        \boldsymbol s_{\mathrm{obs}}
    \right).
\end{equation}
For parameter $z_j$, the predictors consist of the PCA summary and the
preceding transformed parameters $\boldsymbol z_{1:j-1}$, while $z_j$ is the
regression target. Posterior samples are generated sequentially using a fixed
parameter ordering.

The reported NPE-PFN experiments use an ensemble of four TabPFN estimators.
For each conditional predictive distribution, the outer 5\% of each tail is
excluded at the predictive-bucket level to limit extreme autoregressive
samples. A value is then drawn uniformly inside the selected bucket;
additional Gaussian smoothing is disabled. We generate $S=10\,000$
autoregressive posterior samples per run. This internal 5\% truncation is part
of the NPE-PFN sampling procedure and is distinct from the 1\%-trimmed point
summary used only for TSNPE at $B=50$ in the comparison with
$\boldsymbol\theta_{\mathrm{GD}}$.

\paragraph{Posterior summaries.}
For every method, transformed posterior samples are first mapped individually
to physical parameter space,
\begin{equation}
    \boldsymbol\theta^{(r)}
    =
    T^{-1}\!\left(\boldsymbol z^{(r)}\right).
\end{equation}
The posterior mean used for pointwise and output-space evaluation is
\begin{equation}
    \bar{\boldsymbol\theta}
    =
    \frac{1}{S}
    \sum_{r=1}^{S}
    T^{-1}\!\left(\boldsymbol z^{(r)}\right).
\end{equation}
Because $T^{-1}$ is nonlinear, this generally differs from transforming the
mean of the samples in unconstrained space. The only exception is the TSNPE
comparison with $\boldsymbol\theta_{\mathrm{GD}}$ at $B=50$, for which a
componentwise 1\%-trimmed mean is used because an anomalous heavy posterior
tail made the ordinary finite-sample mean unstable. This point is explicitly
marked in Figure~\ref{fig:theta_rmse_physical}; all other method--budget
combinations use the arithmetic posterior mean.

\subsection{Gradient-based calibration details}
\label{app:grad_descente_calib}

The gradient-based reference directly minimizes the discrepancy between the
\texttt{tunax} and LES profiles. All $d_\theta=17$ closure parameters are
optimized jointly in the unconstrained coordinates
$\boldsymbol z=T(\boldsymbol\theta)$, starting from the customary
\texttt{tunax} configuration,
\begin{equation}
    \boldsymbol z_0
    =
    T(\boldsymbol\theta_{\mathrm{default}})
    =
    \boldsymbol\mu_0.
\end{equation}

Unlike the SBI methods, the full-profile calibration does not optimize an
objective in PCA-summary space. Instead, it directly compares the final
simulated and LES profiles:
\begin{equation}
    \mathcal{L}_{\mathrm{full}}(\boldsymbol z)
    =
    \sum_{t\in\mathcal{T}}
    \sum_{f\in\mathcal{F}}
    \frac{1}{|\mathcal{V}_f|}
    \sum_{v\in\mathcal{V}_f}
    \sum_{d\in\mathcal{D}}
    \left[
        \frac{
            \left[
                F\!\left(T^{-1}(\boldsymbol z)\right)
            \right]_{t,f,v,d}
            -
            [\mathbf{x}_{\mathrm{LES}}]_{t,f,v,d}
        }{
            \sigma^{\mathrm{LES}}_{t,f,v}
        }
    \right]^2.
\end{equation}
Here, $\mathcal{T}=\{6,24,72\}$ contains the forecast horizons,
$\mathcal{F}$ the seven forcing regimes,
$\mathcal{V}_f\subseteq\{b,u,v,p_t\}$ the variables retained for forcing $f$,
and $\mathcal{D}$ the 54 calibration depths. For each fixed combination
$(t,f,v)$, $\sigma^{\mathrm{LES}}_{t,f,v}$ is the standard deviation of the
54 corresponding LES values across the depth index $d$. Thus, the same
profile-specific scale normalizes all depthwise residuals in that profile,
preventing variables with larger physical magnitudes from dominating the
loss. The factor $1/|\mathcal{V}_f|$ gives comparable total weight to forcing
regimes containing different numbers of retained variables.

The objective and its gradients are evaluated in 64-bit precision using JAX.
Reverse-mode automatic differentiation propagates through the complete
\texttt{tunax} simulation and turbulence closure. The reported calibration
uses Adam for $1000$ full-batch optimization steps, with a learning rate of
$10^{-2}$ and a maximum gradient norm of 1.

To improve numerical stability and restrict the optimization to a physically
meaningful neighborhood of the customary closure, each transformed parameter
is constrained to
\begin{equation}
    \mu_{0,j}-\log 3
    \leq
    z_j
    \leq
    \mu_{0,j}+\log 3.
\end{equation}
In physical space, this allows the magnitude of each parameter to vary between
one third and three times its default magnitude while preserving its required
sign. Several parameters reach these limits, so the resulting estimate
$\boldsymbol\theta_{\mathrm{GD}}$ should be interpreted as the solution found
within this constrained region rather than as an unconstrained optimum.

Finally, $\boldsymbol\theta_{\mathrm{GD}}$ is a deterministic point estimate
and does not quantify parameter uncertainty. Its computational cost is also
not equivalent to a simulation budget of $1000$ forward evaluations: each
optimization step requires both a forward \texttt{tunax} simulation and
reverse-mode differentiation, making it more expensive than the forward-only simulations used by the SBI methods.


\end{document}